\documentclass[twocolumn,showpacs,
  aps,superscriptaddress,
  prd,notitlepage,showkeys,
  nofootinbib]{revtex4-2}
\usepackage{blindtext}
\usepackage[normalem]{ulem}
\usepackage{amssymb}
\usepackage{amsmath}
\usepackage{graphicx,subfigure}
\usepackage{dcolumn}
\usepackage[colorlinks,urlcolor=blue,citecolor=blue,linkcolor=blue]{hyperref}
\usepackage{color,units}
\usepackage[dvipsnames]{xcolor} 
\usepackage{lineno}
\usepackage{xspace}
\usepackage{longtable} 
\usepackage{float}  
\usepackage{aas_macros}
\usepackage{amsfonts,wasysym,epsfig,verbatim,subfigure,bm,mathrsfs,lipsum}
\usepackage{comment}
\usepackage{braket}
\usepackage{physics}
\usepackage{hyperref}
\usepackage{orcidlink}
\usepackage{tikz}
\usetikzlibrary{positioning}
\usetikzlibrary{fit}
\usetikzlibrary{shapes.geometric, arrows}
\usepackage{placeins}

\tikzstyle{process} = [rectangle, rounded corners, minimum width=2cm, minimum height=1cm, text centered, draw=black, fill=blue!10]
\tikzstyle{arrow} = [thick,->,>=stealth]

\begin{document}

\newcommand{\IUCAA}{Inter-University Centre for Astronomy and
  Astrophysics, Post Bag 4, Ganeshkhind, Pune - 411007, India}

\newcommand{\MPI}{Max-Planck-Institut f{\"u}r Gravitationsphysik (Albert-Einstein-Institut), D-30167 Hannover, Germany}

\newcommand{\LBNZ}{Leibniz Universit{\"a}t Hannover, D-30167 Hannover, Germany}

\newcommand{\WSU}{Department of Physics and Astronomy, Washington State University, 1245 Webster, Pullman, Washington 99164-2814, U.S.A.}

\title{Phenomenological gravitational waveform model of binary black holes \\ incorporating horizon fluxes}

\author{Samanwaya Mukherjee\orcidlink{0000-0002-9055-5784}}
\email{samanwaya@iucaa.in}
\affiliation{\IUCAA}

\author{Khun Sang Phukon \orcidlink{0000-0003-1561-0760}}
\email{k.s.phukon@bham.ac.uk}
\affiliation{School of Physics and Astronomy and Institute for Gravitational Wave Astronomy,\\University of Birmingham, Edgbaston, Birmingham, B15 2TT, United Kingdom}

\author{Sayak Datta}\email{sayak.datta@aei.mpg.de}
\affiliation{\MPI}\affiliation{\LBNZ}

\author{Sukanta Bose}
\affiliation{\WSU}

\begin{abstract}
    
     Subjected to the tidal field of its companion, each component of a coalescing binary suffers a slow change in its mass (tidal heating) and spin (tidal torquing) during the inspiral and merger. For black holes, these changes are associated with their absorption of energy and angular momentum fluxes. This effect modifies the inspiral rate of the binary, and consequently, the phase and amplitude of its gravitational waveform. Numerical relativity (NR) waveforms contain these effects inherently, whereas analytical approximants for the early inspiral phase have to include them manually in the energy balance equation.
     In this work, we construct \texttt{IMRPhenomD\_Horizon}, a frequency-domain gravitational waveform model that incorporates the effects of tidal heating of black holes. This is achieved by recalibrating the inspiral phase of the waveform model \texttt{IMRPhenomD} to incorporate the phase corrections for tidal heating. We also include corrections to the amplitude, but add them directly to the inspiral amplitude model of \texttt{IMRPhenomD}. First we demonstrate that the inclusion of the corrections, especially in the phase, confers an overall improvement in the phase agreement between the analytical inspiral model (uncalibrated \texttt{SEOBNRv2}) and NR data. The model presented here is faithful, with less than 1\% mismatches against a set of hybrid waveforms (except for one outlier that barely breaches this limit). The recalibrated model shows mismatches of up to $\sim 14\%$ with \texttt{IMRPhenomD} for high mass ratios and spins. Amplitude corrections become less significant for higher mass ratios, whereas the phase corrections leave more impact -- suggesting that the former is practically irrelevant for gravitational wave data analysis in Advanced LIGO (aLIGO), Virgo and KAGRA. Comparing with a set of 219 numerical relativity waveforms, we find that the median of mismatches decreases by $\sim 4\%$ in aLIGO zero-detuned high power noise curve, and by $\sim 1.5\%$ with a flat noise curve. This implies a modest but notable improvement in waveform accuracy.

\end{abstract}

\maketitle

\section{Introduction}
\label{sec:intro}

The launch of gravitational wave (GW) astronomy has had a stellar
start, with the detection of over 90 compact binary coalescences (CBCs) so far~\cite{LIGOScientific:2021djp,Nitz:2021zwj,Olsen:2022pin}.
The fourth (O4) observation run of the
ground-based GW detectors LIGO~\cite{LIGOScientific:2014pky}, Virgo~\cite{advVIRGO} and KAGRA~\cite{KAGRA:2018plz}
are expected to detect many more CBCs, which enables us to subject
General Relativity (GR) to unprecedented tests. Such tests demand high
precision CBC waveforms. 
In the field of GW data analysis, waveform models serve as templates against which the real data are compared and estimates of the source parameters are made by the matched filtering technique~\cite{Owen:1998dk} and Bayesian inference~\cite{Veitch:2014wba,Ashton:2018jfp,Romero-Shaw:2020owr}. 
Since a sizable subset of CBCs are high-mass
binaries, which have a good fraction of their signal power arriving
from the late-inspiral and merger parts, these tests benefit from
employing inspiral-merger-ringdown (IMR) waveforms. One needs to solve the full Einstein equations numerically to extract gravitational waveforms predicted by GR to meet the accuracy standard imperative for describing the merger-ringdown parts of a CBC. However, such simulations of numerical relativity (NR) are of extreme computational cost, impeding the usage of this scheme for creating long waveforms that span the entire frequency range of GW detectors for intermediate to stellar-mass binaries. On the other hand, post-Newtonian (PN) approximation (see, e.g., Ref.~\cite{Blanchet:2013haa} and the references therein) or the effective-one-body (EOB) framework~\cite{Buonanno:1998gg,Bohe:2016gbl,Nagar:2018gnk} describes the inspiral part in analytical forms, but fail to be reliably accurate in the late inspiral to the merger regime.  

Efforts to construct accurate IMR models have been based on the EOB formalism with calibrations to NR data (\texttt{SEOBNR}~\cite{Bohe:2016gbl,Taracchini:2012ig,Taracchini:2013rva} and \texttt{TEOBResumS}~\cite{Nagar:2018zoe,Gamba:2021ydi,Nagar:2020pcj} families of models), phenomenological waveform models built by combining PN and NR waveforms~\cite{Ajith:2007kx,Santamaria:2010yb,Hannam:2013oca,Pratten:2020fqn,Husa:2015iqa,Khan:2015jqa}, and more recently, surrogate models~\cite{Tiglio:2021ysj,Field:2013cfa,Blackman:2017pcm}. EOB models generate gravitational waveforms by solving a set of ordinary differential equations. The post-adiabatic (PA) approximation has been used with notable success in the \texttt{TEOBResumS}~\cite{Nagar:2018gnk,Gamba:2020ljo} and some of the \texttt{SEOBNR}~\cite{Mihaylov:2021bpf,Pompili:2023tna} models to solve the dynamical equations analytically, and substantially speeding up the waveform generation. Phenomenological models, in comparison, are equipped with closed-form expressions for the phase and amplitude of the waveforms for a given set of binary parameters. The latter models contain PN-inspired phase and amplitude behaviors augmented with higher-order terms calibrated against a set of ``hybrid waveforms". The hybrids are constructed by stitching either PN (e.g., in \texttt{IMRPhenomC}~\cite{Santamaria:2010yb}) or EOB (e.g. \texttt{IMRPhenomD}~\cite{Khan:2015jqa,Husa:2015iqa}, \texttt{IMRPhenomXAS}~\cite{Pratten:2020fqn} for aligned-spin cases) waveforms with NR ones at a suitable frequency, to allow smooth transitions of phase and amplitude from the inspiral to the merger-ringdown regime. Utmost care needs to be borne when producing complete IMR waveforms by combining these parts, to limit introducing waveform systematics, which can reduce the effectiveness of tests of GR. The increasing modeling errors and computational cost of gravitational waveforms with increasing mass ratio and spins~\cite{Kumar:2016dhh} have inspired the advent of reduced-order models (ROMs) and surrogate models.
By fitting interpolated decomposed waveform data pieces over the binary parameter space, surrogate
models can significantly accelerate NR (e.g.~\cite{Blackman:2015pia,Blackman:2017pcm,Varma:2018mmi}) or EOB waveforms (e.g.~\cite{Purrer:2015tud,Lackey:2016krb,Lackey:2018zvw}), while maintaining high accuracy within their parameter space of validity.

In phenomenological models for binary black holes (BBHs), the inspiral waveform is usually constructed with the ``point-particle" approximation, which also acts as a baseline for creating generic waveforms with tidal effects arising from the finite size of the components, e.g. for binary neutron stars (BNSs)~\cite{Barkett:2015wia,Dietrich:2018uni}. However, even for black holes (BHs) there exists a finite-size effect that needs to be consistently incorporated within a complete BBH waveform model, namely, the absorption of energy and angular momentum by their horizons. Due to their causal structure, BHs in GR are perfect absorbers that behave as dissipative systems~\cite{Poisson:2009di,Cardoso:2012zn}. A significant feature of a BH is its horizon, which is a null surface and a “one-way membrane” that does not allow energy to escape outward. In a binary, a BH experiences tidal interactions from its companion, which cause changes in its mass, angular momentum, and horizon area. This phenomenon is called tidal heating (TH)~\cite{Hartle:1973zz,Poisson:2004cw,Alvi:2001mx,Chatziioannou:2016kem, Saketh:2022xjb, Datta:2023wsn}. This effect is weak but important in many respects. Recent studies~\cite{Datta:2020gem,Mukherjee:2022wws} have found that the presence of TH can be measurable, especially in the future era of GW detectors with orders of magnitude higher signal-to-noise ratios (SNRs) compared to the current ones. Combined with the fact that TH is much more significant for BHs than horizonless compact objects, this makes TH a viable discriminator for horizons to identify BHs against theoretically possible exotic compact objects (ECOs)~\cite{Cardoso:2016oxy}, which can mimic BHs in their GW signals~\cite{Datta:2019epe,Datta:2019euh}.

In NR simulations, TH arises naturally due to the presence of the BH apparent horizons. While evolving the binary, BH singularities are expunged from the computational domain by excision techniques or by the ``moving puncture" method. The properties of the apparent horizons -- the masses, spins, and the horizon areas -- continue to get impacted by the energy and angular momentum fluxes throughout the binary evolution up to the merger. Changes in BH masses and spins impact the inspiral rate of the binary, leaving its imprint in the phase of GW signals. Complete IMR waveforms created with NR data for late inspiral and merger, then, would have incomplete information about BBH systems if they are joined with inspiral approximants devoid of the imprints of horizon fluxes. Ignoring this effect may also lead to ambiguities in the tests of GR from GW observations of CBCs, where small deviations from GR predictions are probed under the strong gravity conditions of a CBC. Lesser systematics will facilitate a more accurate evaluation of GR's validity in such cases.

In the EOB formalism, progress has been made to include the effects of horizon absorption for nonspinning binaries~\cite{Nagar:2011aa,Bernuzzi:2012ku,Damour:2012ky} and for aligned spins~\cite{Damour:2014sva}. These works use numerical evaluation of horizon fluxes using a frequency-domain perturbative approach along stable and unstable orbits in the test-mass limit. This test-mass knowledge is then hybridized with lower-order analytical information that is valid for comparable masses. These calculations were implemented in the construction of \texttt{TEOBResumS}~\cite{Nagar:2018zoe}, an NR-informed EOB waveform model which includes the horizon-absorption effects in a PN-inspired resummed form.  Alvi~\cite{Alvi:2001mx} has calculated, under PN approximation, leading-order expressions of mass and spin evolution of BBH systems. Alvi's expressions are valid for binaries with comparable masses. A resummed version of these expressions was used by Damour and Nagar~\cite{Damour:2014sva} to describe them under the EOB framework.

In Phenomenological models, horizon absorption was taken into account in the flux calculations of \texttt{IMRPhenomC}~\cite{Santamaria:2010yb}, up to 2.5PN order. However, later models with more accurate calibrations spanning larger regions of the parameter space (e.g. \texttt{IMRPhenomD} or \texttt{IMRPhenomXAS}) do not account for this effect explicitly in their inspiral parts. 

In this work, we construct a phenomenological BBH waveform model with a consistent inclusion of TH effects, and study its contribution to an improvement in waveform systematics. A complete BBH model, which includes the phase and amplitude modifications for TH from the early inspiral to the merger, not only provides a comprehensive foundation but also helps to lay the groundwork for a more general waveform model with the \textit{horizon parameters}, which can act as discriminators for BHs from ECOs or NSs~\cite{Datta:2020gem,Mukherjee:2022wws}.

We first describe some preliminaries and notations of gravitational waveforms in Sec.~\ref{sec:convention}. Then in Sec.~\ref{sec:inspiral}, we build an inspiral model containing explicit corrections due to TH at 2.5PN, 3.5PN, and 4PN orders in their phase and amplitude. The point-particle baseline for the inspiral waveform is based on an aligned-spin EOB approximant. In Sec.~\ref{sec:NR}, we describe the publicly available NR waveforms of the SXS catalog~\cite{Boyle:2019kee} used for merger and ringdown phases. We investigate the phase agreement between NR data and the analytical inspiral waveforms in Sec.~\ref{sec:phase comparison}. In Sec.~\ref{sec:hyb} we describe the construction of new hybrid waveforms by stitching together the TH-corrected inspiral and merger-ringdown parts. Following \texttt{IMRPhenomD} (henceforth referred to as \texttt{PhenomD}), we model the hybrids in the frequency domain in Sec.~\ref{sec:model} and calibrate the inspiral part with augmented PN expressions, but against the new set of hybrid waveforms. Section~\ref{sec:model validation} discusses the faithfulness of the new model, and in Sec.~\ref{sec:comparison}, we compare it with \texttt{PhenomD} and NR data within the parameter range of calibration. We conclude in Sec.~\ref{sec:discussion} with discussions and future directions of this work.

Throughout the article, we will use geometric units by considering the gravitational constant ($G$) and the speed of light in vacuum ($c$) set to unity, except while calculating physical quantities.

\section{Waveform conventions and notation}\label{sec:convention}
We consider $l=|m|=2$ spherical harmonic modes of gravitational waves from coalescing binary black holes. In the time domain, the complex GW strain can be written as,
\begin{equation}\label{eq:strain}
    h^{}_{2,2}(\Theta,t)=A(\Theta,t)e^{-i\phi(\Theta,t)},
\end{equation}
where $\Theta$ is the parameter vector containing the parameters of the binary. The intrinsic parameters are the dimensionless spin vectors $\vb*{\chi^{}_1},\vb*{\chi^{}_2}$ and the two masses $m^{}_1$ and $m^{}_2$. Since we are concerned with non-precessing binaries only, the spin vectors have nonzero components only along (or opposite to) the direction of the orbital angular momentum. In terms of the normalized orbital angular momentum $\hat{L}$ and the dimensionful spin vectors $\vb*{S_i}$ ($i=1,2$ for the two  objects), the components of dimensionless spins can be expressed as
\begin{equation}\label{eq:spins_def}
    \chi^{}_i=\frac{\vb*{S^{}_i}\cdot \hat{L}}{m_i^2}\,.
\end{equation}
We define the mass ratio as $q=m^{}_1/m^{}_2\geqslant 1$, total mass $M=m^{}_1+m^{}_2$, and the symmetric mass ratio $\eta=m^{}_1m^{}_2/M^2$. 

The complex strain in Eqn.~\eqref{eq:strain} can be constructed from the plus and cross polarization states as
\begin{equation}
    h(\Theta,t)=h_{+}(\Theta,t)-ih_{\cross}(\Theta,t)\,.
\end{equation}
The Fourier transform of the complex strain is defined as
\begin{equation}
    \Tilde{h}(f)=\int_{-\infty}^\infty h(t)e^{-i2\pi ft}\dd t\,,
\end{equation}
where $f$ denotes frequency. For the non-eccentric waveforms that we use, $f$ can be written as a monotonically increasing function of time $t$ until the merger. 

In gravitational waveform modeling, the dimensionless frequency $Mf$ plays an important role, since the total mass $M$ acts as a scaling factor. For better readability, we define $F\equiv Mf$.

\section{Input waveforms}

\subsection{Inspiral waveforms}\label{sec:inspiral}
\subsubsection{Effective-one-body description for the point-particle waveforms}

In the effective-one-body approach to the two-body problem in GR, originally developed by Buonanno and Damour~\cite{Buonanno:1998gg}, the dynamics of two compact objects of masses $m^{}_1$ and $m^{}_2$ and spins $\vb*{\chi^{}_1}$ and $\vb*{\chi^{}_2}$ is mapped onto the dynamics of an effective particle of mass $\mu = m^{}_1m^{}_2/(m^{}_1+m^{}_2)$ and spin $\vb*{\chi^{}_\ast}$ moving in the background of a deformed Kerr geometry with mass $M=m^{}_1+m^{}_2$ and spin $\vb*{\chi}^{}_{\rm Kerr}$. The spin mapping $\{\vb*{\chi^{}_1},\vb*{\chi^{}_2}\}\to \vb*{\chi^{}_\ast}$ and the deformation of the Kerr background, parametrized by the symmetric mass ratio $\eta$, implies that the test-particle dynamics reproduces the PN-expanded dynamics of the original two-body system. The two primary EOB families of waveforms, namely the \texttt{SEOBNR} and \texttt{TEOBResumS} families, differ in the ways they describe the conservative Hamiltonian, the amount of PN and NR information they utilize, and the radiative parts of the waveforms~\cite{Rettegno:2019tzh}.

For the purpose of creating hybrid waveforms in our work, we choose \texttt{SEOBNRv2}~\cite{Taracchini:2013rva}, an aligned-spin EOB approximant, as the point-particle baseline. More recent versions of the \texttt{SEOBNR} family of waveforms are available~\cite{Bohe:2016gbl,Pompili:2023tna} at present, but we create our model based on the former one for a direct comparison with \texttt{PhenomD}. 
We remove the NR calibrations present in the \texttt{SEOBNRv2} model, following the construction of \texttt{PhenomD}, since fitting to NR data may leave residues of tidal heating imprints at higher frequencies. To remove the NR calibrations, we perform the changes in the \texttt{SEOBNRv2} model in \texttt{LALSuite}~\cite{lalsuite} as mentioned in Sec. II C in Ref.~\cite{Husa:2015iqa}. The uncalibrated version of \texttt{SEOBNRv2}, henceforth referred to as \texttt{SEOBv2}, serves as the point-particle baseline for creating the hybrid waveforms. 
To incorporate the corrections due to TH in BBH systems, we transform this time-domain waveform to the frequency domain to add the phase and amplitude corrections, as described in the next section.

\subsubsection{Corrections for tidal heating of black holes}

The early inspiral part of a CBC can be described by the analytical formalism of post-Newtonian (PN) expansions~\cite{Blanchet:2013haa}, under the approximation that the source is slowly moving and weakly gravitating. In this formalism, the evolution of the orbital phase $\Psi(t)$ of a compact binary is computed as a perturbative expansion in a small parameter, typically taken to be the characteristic velocity $v = (\pi M f)^{1/3}$. This analytical procedure demands $v\ll 1$, which makes it useful in the early inspiral phase of a CBC.

An electrically neutral spinning black hole in GR -- the Kerr black hole (KBH) -- is stationary when it is isolated. On the other hand, when a KBH is a member of a binary, it feels its companion's tidal field, which acts as a non-axisymmetric perturbation~\cite{Hartle:1973zz}. This perturbation causes changes in the mass, spin, and horizon area of the KBH over time~\cite{Alvi:2001mx}. Since the KBH experiences the tidal field of its orbiting companion, it absorbs (emits) energy from (into) the orbit. The absorption part is present in non-spinning BHs as well. Additionally, for a KBH, a spin frequency higher than the angular frequency of the tidal field slows down due to tidal interactions,  which in turn makes the KBH lose its rotational energy. The slowing down of a rotating BH due to the gravitational dissipation produced by exterior mass is analogous to the slowing down of a rotating planet by viscous dissipation due to tides raised by an exterior moon that increases its internal thermal content - a phenomenon known as tidal heating. Due to this similarity, the energy and angular momentum flux in BBHs is also termed tidal heating~\cite{Poisson:2004cw}.

During the early stages of the binary evolution, the change in the orbital frequency due to emission of GWs is much smaller than the orbital frequency itself. In this \textit{adiabatic inspiral} regime, loss of binding energy $E(v)$ of the two-body system with time equals the GW flux emitted to future null infinity ($\mathcal{F}^{\infty}(v)$) plus the energy flux absorbed by the two BH horizons ($\mathcal{F}^{\rm H}(v)$). So the energy balance condition becomes
\begin{equation}
    -\dv{E(v)}{t}=\mathcal{F}^{\infty}(v)+\mathcal{F}^{\rm H}(v).
\end{equation}

Evolution of the orbital phase $\phi$ and the characteristic velocity $v$, obtained from this equation, read
\begin{equation}
    \dv{\phi}{t} = \frac{v^3}{M}\,,\quad\quad\quad \dv{v}{t}=-\frac{\mathcal{F}(v)}{E'(v)},
\end{equation}
where $\mathcal{F}(v)=\mathcal{F}^{\infty}(v)+\mathcal{F}^{\rm H}(v)$. These equations yield a solution for the phase $\Psi(f)$ of the frequency-domain waveform $\Tilde{h}(f) =  \Tilde{A}(f) e^{-i\Psi(f)}$~\cite{Tichy:1999pv}:
\begin{equation}
\label{psi}
    \Psi(f)=2(t_c/M)v^3 - 2\phi_c - \pi/4 - \frac{2}{M}\int (v^3-\Bar{v}^3)\frac{E'(\Bar{v})}{\mathcal{F}(\Bar{v})}\dd\Bar{v},
\end{equation}
where $E'(v)=\dv*{E(v)}{v}$. 

The frequency-domain amplitude $\Tilde{A}(f)$ can be written as\footnote{Note that we define $E(v)$ as the total binding energy. Ref~\cite{Ajith:2011ec} treats $E(v)$ as the specific binding energy (binding energy per unit mass), so the power of $M$ in our treatment is different from theirs in this particular equation.}~\cite{Ajith:2011ec}
\begin{equation}
\label{amp}
    \Tilde{A}(f)=\mathcal{C}\frac{2\eta M^{3/2}}{d^{}_L}v\sqrt{\frac{\pi}{3}}\left[-\frac{E'(v)}{\mathcal{F}(v)}\right]^{1/2},
\end{equation}
where $\mathcal{C}$ is a constant that depends on the inclination of the orbital plane with the line of sight, and $d^{}_L$ denotes the distance to the binary.

Splitting $\mathcal{F}$ into $\mathcal{F}^\infty$ and $\mathcal{F}^{\rm H}$ in Eq.~\eqref{psi} enables one to write the frequency-domain phase in the form
\begin{equation}
\label{eq:psi2}
    \Psi(f)=\Psi^{}_{\rm PP}(f)+\Psi^{}_{\rm TH}(f),
\end{equation}
with $\Psi^{}_{\rm PP}(f)$ being the GW phase under point particle (PP) approximation that corresponds to $\mathcal{F}^\infty$, and $\Psi^{}_{\rm TH}(f)$ being the phase correction due to TH.

The most recent expression for the dephasing $\Psi^{}_{\rm TH}(f)$ is based on the effective worldline theory in Ref.~\cite{Saketh:2022xjb}, and it matches with the results in the extreme mass ratio limit~\cite{Tagoshi:1997jy}. In this work, we use the aforementioned results to calibrate our BBH waveform model with tidal heating. Up to 4PN order, $\Psi^{}_{\rm TH}(f)$, or equivalently $\Psi^{}_{\rm TH}(v)$, can be expressed as (for the (2,2) mode)
\begin{equation}\label{eq:phase correction}
    \Psi^{}_{\rm TH}(v) = \frac{3}{128\eta v^5}\sum_{n=5}^8 \left[ \delta\psi^{}_{(n)} v^n + \delta\psi^{}_{(n,l)}v^n \log v\right]\,.
\end{equation}
Explicit expressions for $\delta\psi^{}_{(n)}$ and $\delta\psi^{}_{(n,l)}$ are given in Sec. V of Ref~\cite{Saketh:2022xjb}.

Amplitude corrections for TH in the frequency domain have been derived for the first time in the current work. We find the PN expansion of the total amplitude by expanding $\left[-E'(v)/\mathcal{F}(v)\right]^{1/2}$ in Eq.~\eqref{amp} in powers of $v$ up to $v^7$, and separate the contribution due to TH:
\begin{equation}\label{eq:af}
    \Tilde{A}(f)=\Tilde{A}^{}_{\rm PP}(f)+\Tilde{A}^{}_{\rm TH}(f).
\end{equation}
 
$\Tilde{A}^{}_{\rm PP}(f)$ is mentioned in Eq.~(5.7) of Ref.~\cite{Ajith:2011ec}, and we get the expression for $\Tilde{A}^{}_{\rm TH}(f)$ as
\begin{equation}\label{eq:amp correction}
\begin{aligned}
    \Tilde{A}^{}_{\rm TH}(f)= ~ \mathcal{C}&\frac{M^{5/6}}{d^{}_L\pi^{2/3}}\sqrt{\frac{5\eta}{24}}f^{-7/6}\left[\frac{1}{8}H^{(\rm BBH)}_{\rm eff5}v^5 \right.\\ 
    &\left.~+\left(\frac{1079}{1792}+\frac{103}{192}\eta\right)H^{(\rm BBH)}_{\rm eff5}v^7 \right. \\ &\left. + \frac{1}{32}\left\{\Tilde{A}^{}_{\rm SO}H^{(\rm BBH)}_{\rm eff5} \right.\right.\\ &\left.\left. - 8\left(H^{(\rm BBH)}_{\rm eff8}-\pi H^{(\rm BBH)}_{\rm eff5}\right)\right\} v^8\right],
\end{aligned}
\end{equation}
where the quantities $H^{(\rm BBH)}_{\rm eff5}$ and $H^{(\rm BBH)}_{\rm eff8}$ are defined as
\begin{subequations}
\label{Eq.Hparams}
\begin{align}
H^{(\rm BBH)}_{\rm eff5} \equiv &{} \sum_{i=1}^{2} \left(\frac{m^{}_i}{M}\right)^3 \left(\hat{L}\cdot\hat{S}^{}_i\right)\chi^{} _i \left(3 \chi^{}_i{}^2+1\right)\,,\\
H^{(\rm BBH)}_{\rm eff8} \equiv &{} ~4 \pi  H^{(\rm BBH)}_{\rm eff5}+\sum^2_{i=1} \left(\frac{m_i}{M}\right)^4 \left(3 \chi^{}_i{}^2+1\right)\nonumber \\
                &\quad\quad\quad\quad\quad\quad\quad \times \left(\sqrt{1-\chi^{}_i{}^2}+1\right)\,.
\end{align}
\end{subequations}

$\Tilde{A}^{}_{\rm SO}$ is the spin-orbit term 
\begin{equation}
\begin{aligned}  
    \Tilde{A}^{}_{\rm SO} = &~ \frac{179}{6(1+q)^2}\left\{q^2(\hat{L}\cdot\hat{S}^{}_1)\chi^{}_1 + (\hat{L}\cdot\hat{S}^{}_2)\chi^{}_2\right\} \\ & 
                    +\frac{35q}{2(1+q)^2}\left\{(\hat{L}\cdot\hat{S}^{}_1)\chi^{}_1 + (\hat{L}\cdot\hat{S}^{}_2)\chi^{}_2\right\}.
\end{aligned}
\end{equation}

\texttt{SEOBv2} is a time-domain waveform, and the phase and amplitude corrections for TH are calculated in the frequency domain. So, we have to first convert the time-domain data to the Fourier domain to apply the corrections. Since the time-domain data is finite, we use a tapered window function to minimize Gibbs artifacts, called the Planck-taper window~\cite{MacDonald:2011ne}:
        
        \begin{equation}
            \begin{aligned}
                w(x) =
            \left\{
        	    \begin{array}{ll}
        		    0 , &  x \leqslant x^{}_1 \\
        		    \left[e^{y(x)}+1\right]^{-1}, &  x^{}_1 < x < x^{}_2 \\
        		    1 , &  x^{}_2 \leqslant x \leqslant x^{}_3 \\
        		    \left[e^{z(x)}+1\right]^{-1}, &  x^{}_3 < x < x^{}_4 \\
        		    0, &  x^{}_4 \leqslant x,
        	    \end{array}
            \right.
            \end{aligned}
        \end{equation}
        where $y(x) = (x^{}_2-x^{}_1)/(x-x^{}_1)+(x^{}_2-x^{}_1)/(x-x^{}_2)$, and $z(x)=(x^{}_3-x^{}_4)/(x-x^{}_3)+(x^{}_3-x^{}_4)/(x-x^{}_4)$. 

        The frequency-domain waveform, after correcting for TH, is then converted back to the time domain. For ease of reference within this paper, we call the resulting waveform \texttt{SEOBv2\_Horizon}. In Appendix~\ref{app: wf models} we list the waveform models used in this work for various purposes.


\subsection{Numerical relativity waveforms for merger-ringdown}\label{sec:NR}

As a BBH progresses towards the merger phase, the BHs come so close to each other that the system becomes too compact for the weak-gravity condition to hold. In addition, their orbital velocities culminate to values comparable to $c$, breaching the slow-motion approximation. The analytical PN results, with their highest order of expansion available currently, become increasingly inaccurate to describe the system evolving under such extreme conditions. 
Starting from this \textit{late inspiral} phase to the merger and ringdown phase where the two BHs merge and the final remnant BH settles to equilibrium, one needs to solve the full Einstein equations numerically for extracting GWs to conform to the accuracy standards of the waveforms required for searches of GW signals or their parameter estimation (PE). This is the field of NR, which has seen exciting breakthroughs ~\cite{Pretorius:2005gq,Campanelli:2005dd,Baker:2005vv} in the early 2000s, and great strides towards simulating merging black holes with different masses and spins~\cite{Pfeiffer:2012pc,Centrella:2010mx,LeTiec:2014oez,Hannam:2013pra}.

We use NR data from the publicly available SXS catalog~\cite{Boyle:2019kee}, computed using the Spectral Einstein Code (SpEC)~\cite{Scheel:2008rj,Ossokine:2013zga}. SpEC uses the excision techniques to remove the BH singularity for extracting gravitational waves from binary black hole systems. Unlike the PN framework where the fluxes of energy and angular momentum absorbed by the BHs have to be considered explicitly in the energy balance equation, numerical simulations of full Einstein equations capture this effect inherently. Scheel \textit{et al.}~\cite{Scheel:2014ina} have demonstrated the slow change in the BH masses and spins through the binaries' temporal evolution in NR simulations with SpEC. They track the apparent horizons as a function of time, and at frequent time intervals, they measure both the surface area and the spin of the horizons. The spin computation has been carried out using the approximate Killing vector formalism~\cite{Cook:2007wr}. The mass of the black hole is then computed using Christodoulou's formula~\cite{Christodoulou:1971pcn},
\begin{equation}
    M^2=M_{\rm irr}^2 + \frac{S^2}{4M_{\rm irr}^2},
\end{equation}
where $M_{\rm irr}$ is the irreducible mass of the Kerr BH,
\begin{equation}
    M_{\rm irr}^2=\frac{1}{16\pi}\int_\mathcal{H}\dd A.
\end{equation}
Here $S$ is the (dimensionful) spin of the KBH and the integration covers the total horizon area $A$.

During the evolution, each SXS NR waveform is extracted at a series of times on a set of concentric coordinate spheres surrounding the binary, decomposed in modes of spin-weighted spherical harmonic functions. Then the waveforms are extrapolated to future null infinity $\mathscr{I}^+$. The dataset of each SXS waveform contains several gravitational waveform modes $(l,m)$, and the orders of extrapolation $N=2,3,4$. A higher order of extrapolation is preferred when accurate waveforms are needed in the inspiral, and lower order extrapolation is preferred for more accuracy at the merger-ringdown phase~\cite{Boyle:2019kee}. Since we use these waveforms for constructing hybrids, we need more accuracy for the merger-ringdown part, due to the fact that some part of the inspiral has to be replaced by the PN waveforms in the hybrids. We use 20 aligned-spin non-eccentric BBH waveforms from SXS with $1\leqslant q \leqslant 8$ and $-0.95\leqslant \chi^{}_1,\chi^{}_2\leqslant 0.85$ for calibrating our model. We choose $(2,2)$ modes, and the extrapolation order $N=2$ for our purpose.

\subsection{Phase comparison of inspiral and NR waveforms}\label{sec:phase comparison}

\begin{figure}
\centering
 \subfigure[]{\label{fig:phasediff1a}\includegraphics[width=\linewidth]{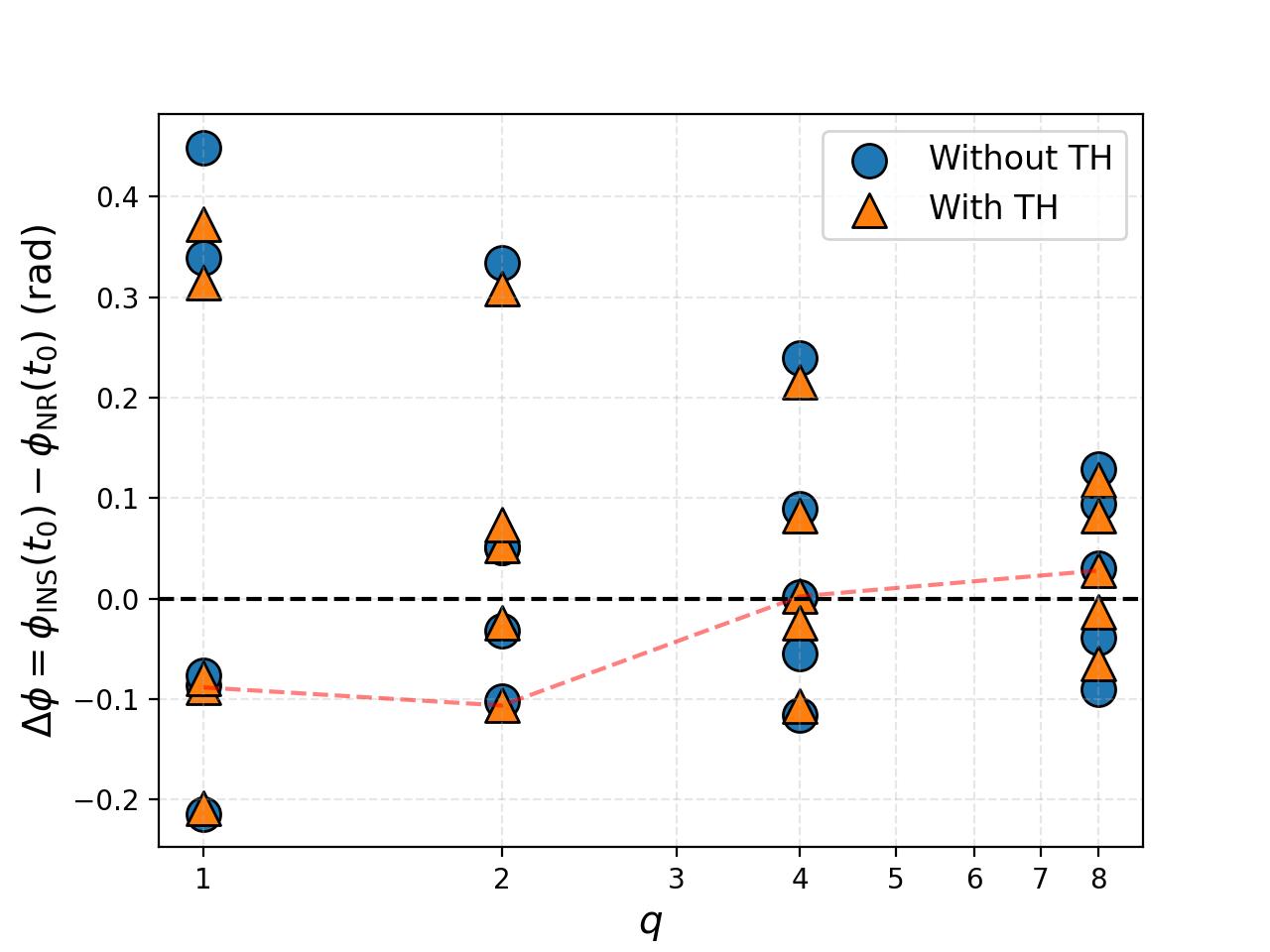}}
 \hfill
 \subfigure[]{\label{fig:phasediff1b}\includegraphics[width=\linewidth]{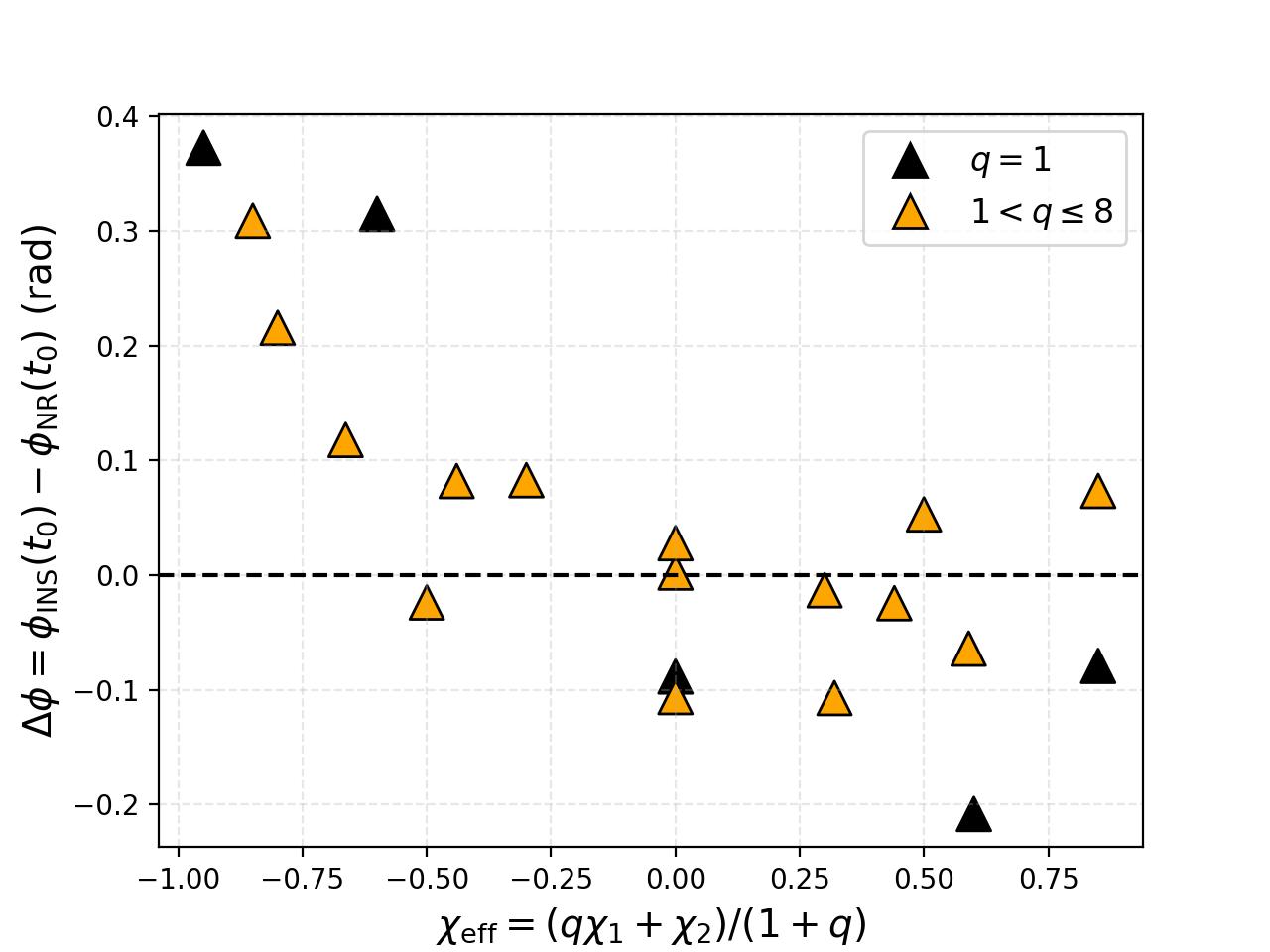}}
 \caption{Accumulated phase difference between the inspiral and NR waveforms in 10 GW cycles of the latter prior to $M\omega^{}_{\rm GW}=0.075$: $\phi^{}_{\rm NR}(0)-\phi^{}_{\rm NR}(t_0)=20\pi$. The NR data correspond to the 20 NR waveforms used for hybrid construction, described in Table~\ref{tab:hyb}. (a) Top panel: variation with the mass ratio $q$, shown for two cases of the inspiral waveforms (\texttt{SEOBv2}), one with the tidal heating corrections and one without. The red dashed line connects the nonspinning binaries. (b) Bottom panel: same points (with the tidal heating corrections included) as in the top panel, but plotted against the effective spin parameter $\chi^{}_{\rm eff}=(q\chi^{}_1+\chi^{}_2)/(1+q)$. Equal-mass binaries are highlighted.}
\label{fig:phasediff}
\end{figure}

\begin{figure}
\centering
\subfigure[]{\label{fig:phasediff2a}\includegraphics[width=0.85\linewidth]{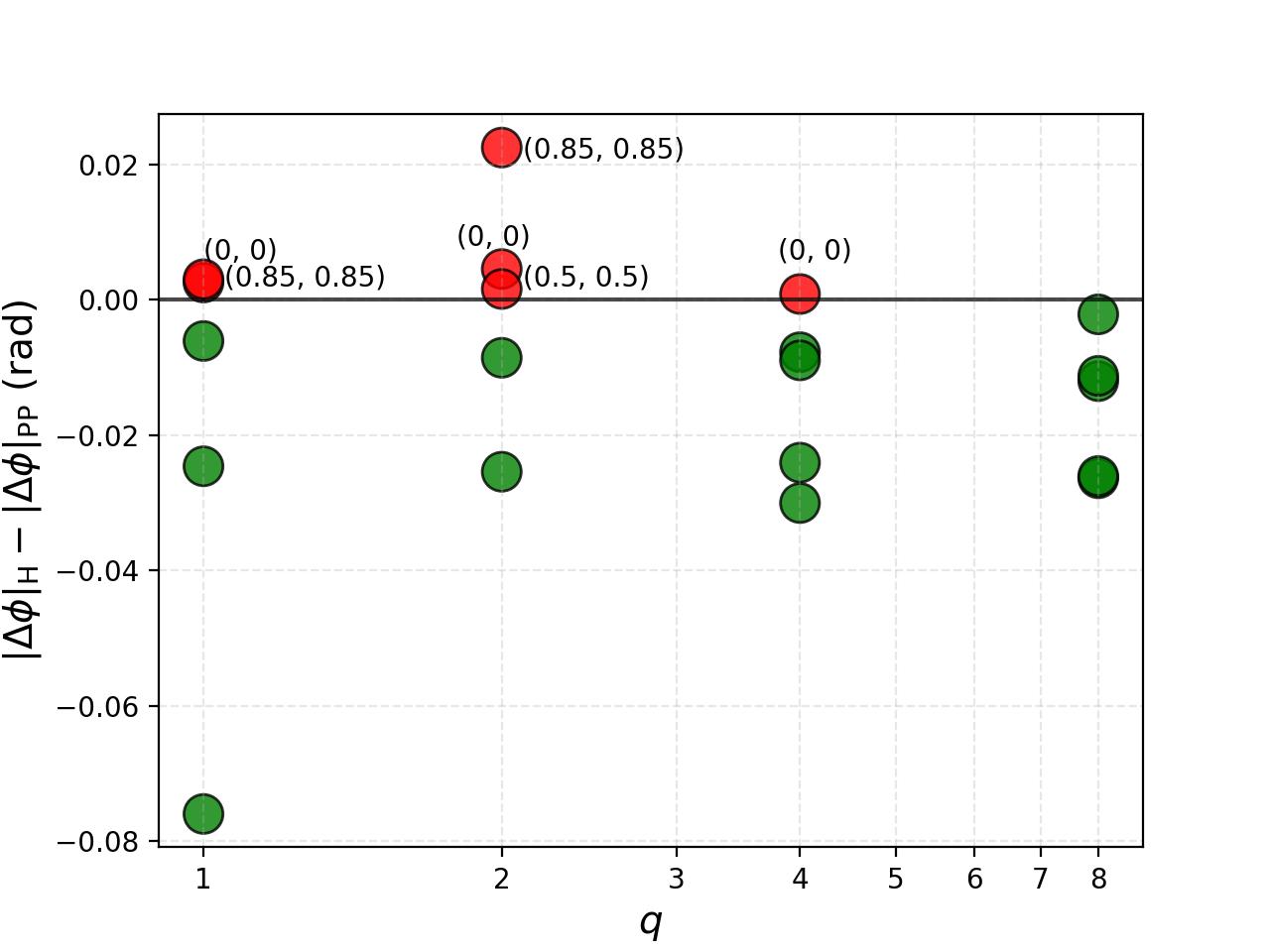}}
 \hfill
 \subfigure[]{\label{fig:phasediff2b}\includegraphics[width=0.85\linewidth]{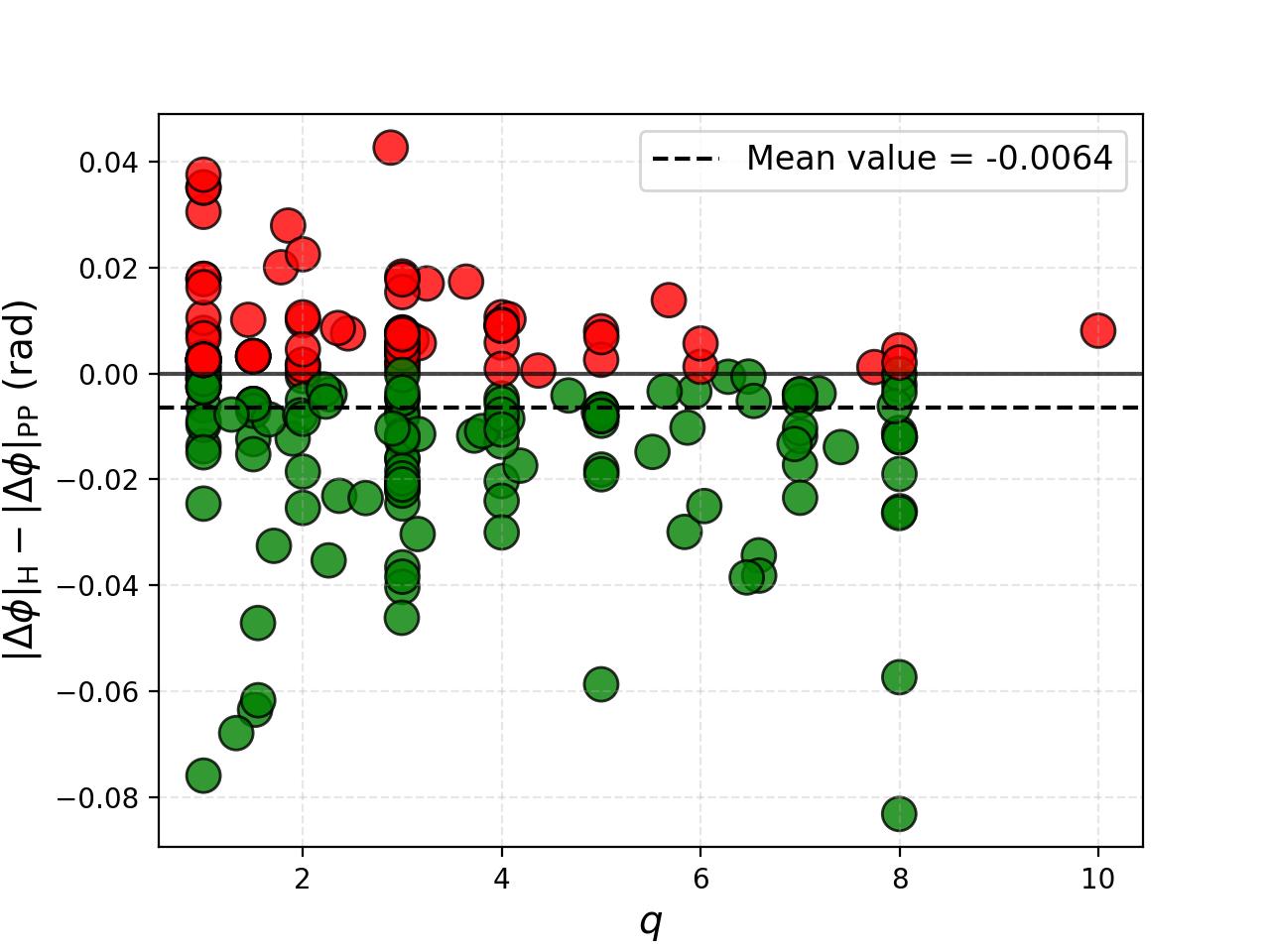}}
 \hfill
 \subfigure[]{\label{fig:phasediff2c}\includegraphics[width=0.85\linewidth]{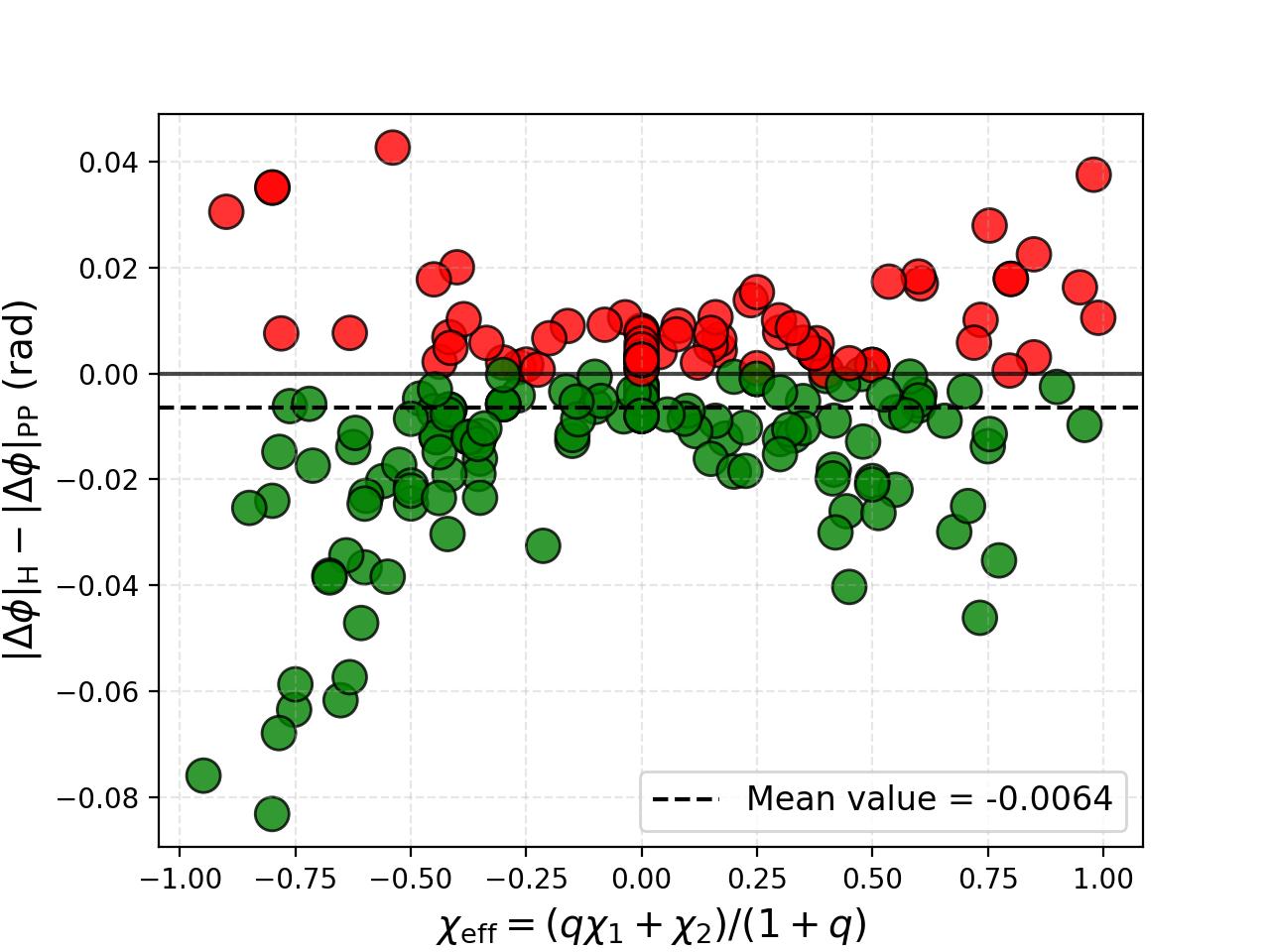}}
 \caption{Difference between the accumulated phase errors (Figure~\ref{fig:phasediff}) with and without the tidal heating corrections. (a) Top panel: the 20 NR waveforms used in Fig.~\ref{fig:phasediff}. The subscript H (PP) refers to the \texttt{SEOBv2} model with (without) the TH corrections. Spin values are reported for the binaries that produce higher phase errors compared to NR after adding the TH corrections ($|\Delta\phi|^{}_{\rm H}-|\Delta\phi|^{}_{\rm PP}\geq 0$), plotted in red color. Binaries with lesser phase disagreement after the corrections are plotted in green. (b) Middle panel: same as in the top panel, with 218 NR waveforms from the SXS catalog, plotted against the mass ratio. (c) Bottom panel: same as in the middle panel, plotted against the effective spin parameter.}
\label{fig:phasediffNR}
\end{figure}

\begin{table*}
    \centering
    \begin{tabular}{p{0.05\linewidth}p{0.18\linewidth}p{0.08\linewidth}p{0.08\linewidth}p{0.08\linewidth}p{0.08\linewidth}p{0.08\linewidth}p{0.08\linewidth}}
        \hline
        \# & Simulation label & $q$ & $\chi^{}_1$ & $\chi^{}_2$ & $\chi^{}_{\rm PN}$ & $N_{\rm orb}$ & $F_{\rm stitch}$\\
        \vspace{0.5mm}\\
        \hline
        \hline
        \vspace{0.5mm}\\
        1 & SXS:BBH:0156 & 1 & $-$0.95 & $-$0.95 & $-$0.79 & 13 & 0.0057 \\
        2 & SXS:BBH:0151 & 1 & $-$0.6 & $-$0.6 & $-$0.5 & 15 & 0.0054 \\
        3 & SXS:BBH:0001 & 1 & 0 & 0 & 0 & 28 & 0.0041 \\
        4 & SXS:BBH:0152 & 1 & 0.6 & 0.6 & 0.5 & 23 & 0.0052 \\
        5 & SXS:BBH:0153 & 1 & 0.85 & 0.85 & 0.7 & 25 & 0.0052 \\
        6 & SXS:BBH:0234 & 2 & $-$0.85 & $-$0.85 & $-$0.72 & 28 & 0.0037 \\
        7 & SXS:BBH:0238 & 2 & $-$0.5 & $-$0.5 & $-$0.43 & 32 & 0.0037 \\
        8 & SXS:BBH:0169 & 2 & 0 & 0 & 0 & 16 & 0.0062 \\
        9 & SXS:BBH:0253 & 2 & 0.5 & 0.5 & 0.43 & 29 & 0.0046 \\
        10 & SXS:BBH:2131 & 2 & 0.85 & 0.85 & 0.72 & 25 & 0.0056 \\
        11 & SXS:BBH:1936 & 4 & $-$0.8 & $-$0.8 & $-$0.71 & 17 & 0.0057 \\
        12 & SXS:BBH:1418 & 4 & $-$0.4 & $-$0.5 & $-$0.37 & 67 & 0.005 \\
        13 & SXS:BBH:0167 & 4 & 0 & 0 & 0 & 16 & 0.0071 \\
        14 & SXS:BBH:1417 & 4 & 0.4 & 0.5 & 0.37 & 80 & 0.0057 \\
        15 & SXS:BBH:1907 & 4 & 0 & 0.8 & 0.12 & 21 & 0.0062 \\
        16 & SXS:BBH:1423 & 8 & $-$0.6 & $-$0.75 & $-$0.57 & 18 & 0.0066 \\
        17 & SXS:BBH:0064 & 8 & $-$0.5 & 0 & $-$0.43 & 19 & 0.0066 \\
        18 & SXS:BBH:0063 & 8 & 0 & 0 & 0 & 26 & 0.0065 \\
        19 & SXS:BBH:0065 & 8 & 0.5 & 0 & 0.43 & 34 & 0.0063 \\
        20 & SXS:BBH:1426 & 8 & 0.48 & 0.75 & 0.47 & 26 & 0.0078 \\
        \vspace{0.5mm}\\
        \hline
    
    \end{tabular}
    \caption{Hybrid waveforms used to calibrate the model. The first column lists the simulation IDs of the corresponding NR data in the SXS catalog. The last column reports the dimensionless frequency at the midpoint of the stitching region.}
    \label{tab:hyb}
\end{table*}

\begin{figure*}
\centering
 \subfigure[]{\includegraphics[width=0.49\textwidth]{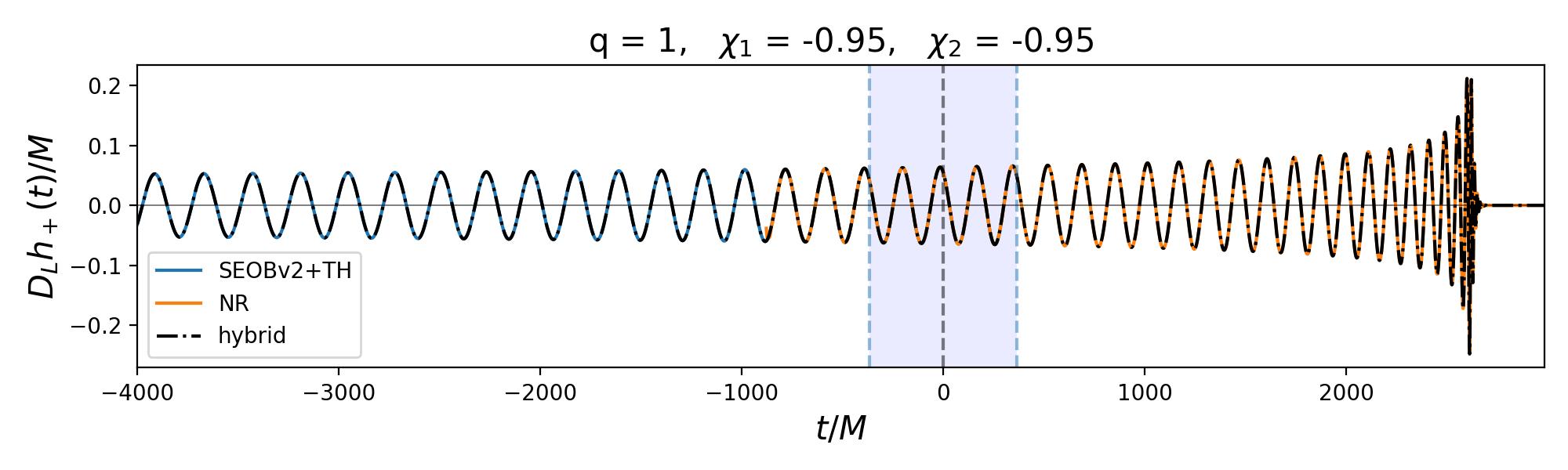}}
 \hfill
 \subfigure[]{\includegraphics[width=0.49\textwidth]{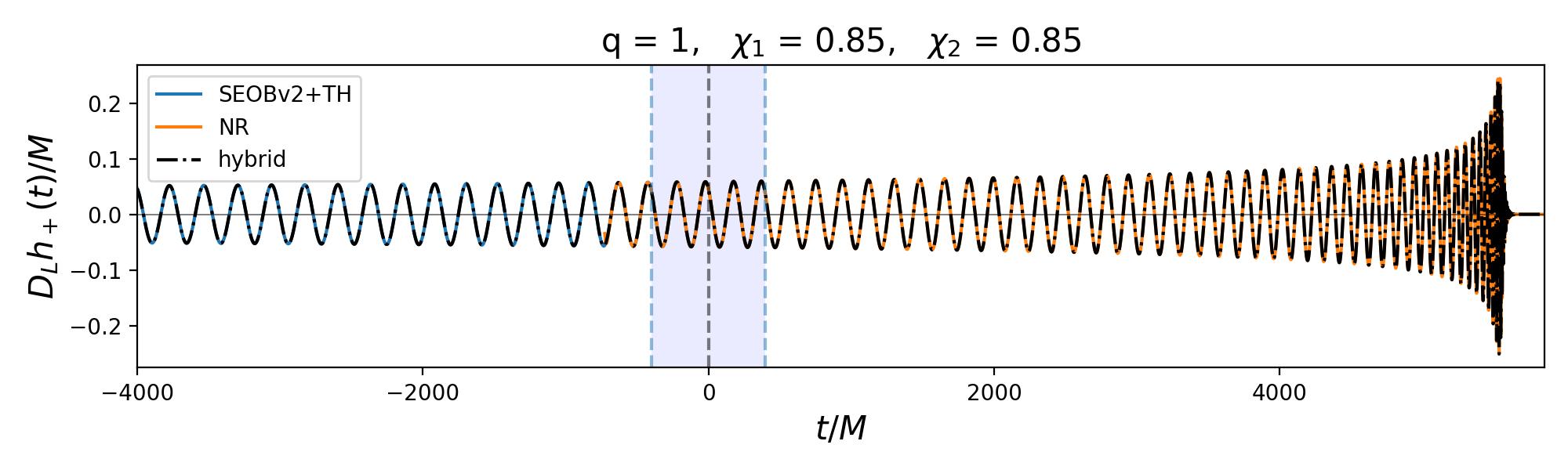}}
 \hfill
 \subfigure[]{\includegraphics[width=0.49\textwidth]{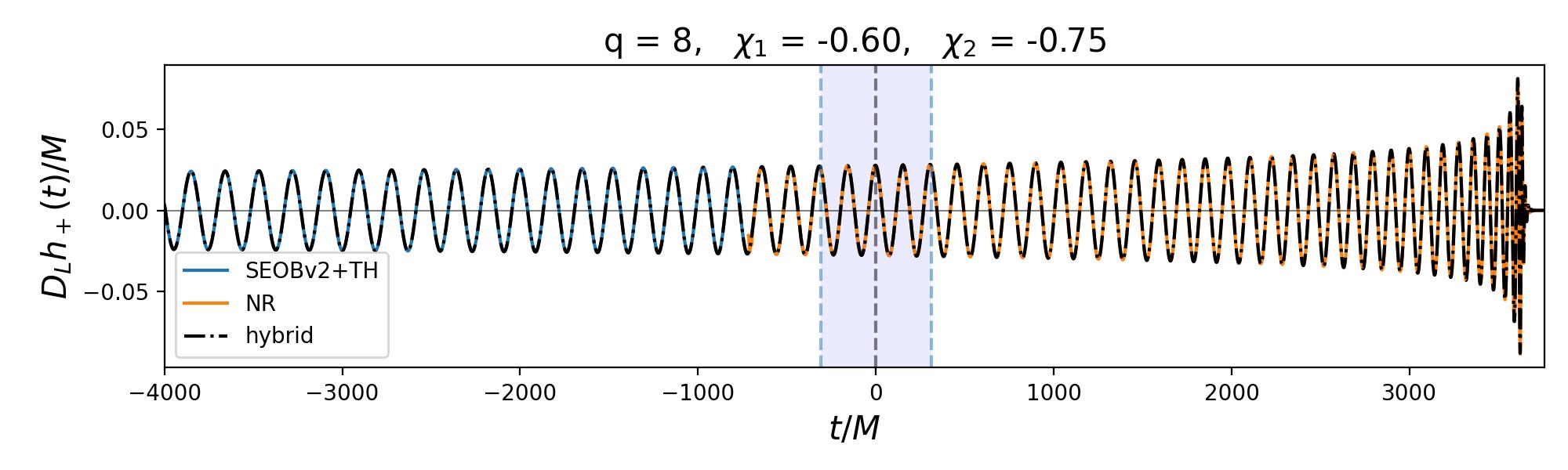}}
 \hfill
 \subfigure[]{\includegraphics[width=0.49\textwidth]{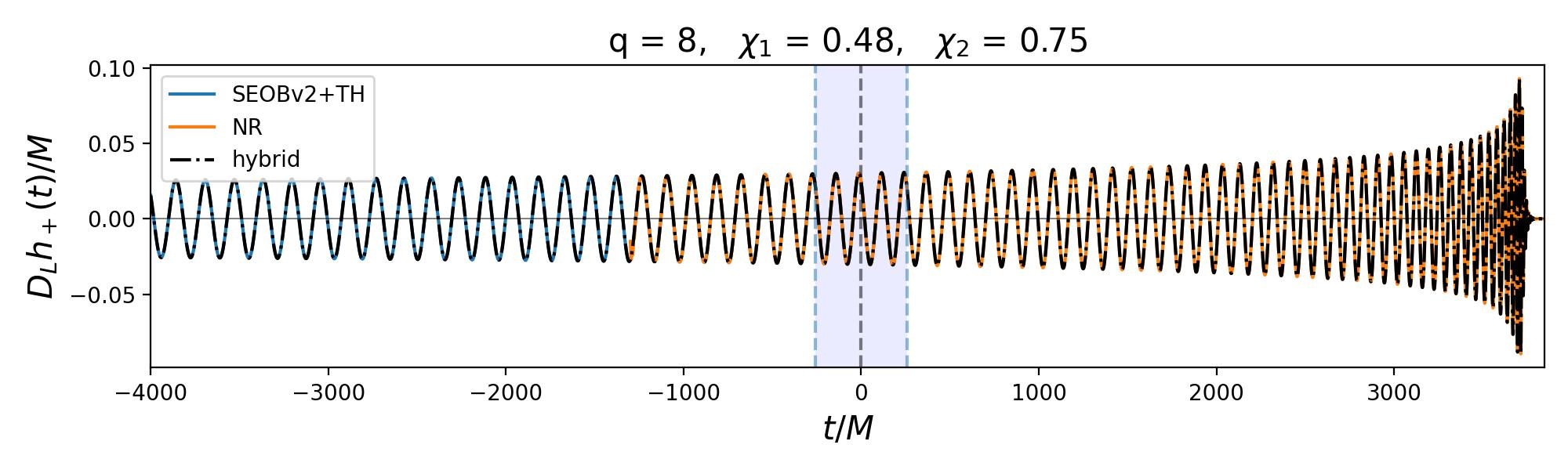}}
 \caption{Hybrid waveforms for four different configurations in the parameter space of $q,\chi^{}_1,\chi^{}_2.$ Inspiral waveforms are generated by adding the phase and amplitude corrections due to TH to the \texttt{SEOBv2} model, shown in blue. NR waveforms are shown in orange, and the hybrid waveforms are shown in black dashed-dotted lines. The $x$ and $y$ axes denote time (in units of total mass) and the real part of time-domain strain, $h_+(t)$ (in units of $M/D_L$), respectively. The blue shaded areas denote the stitching regions.}
\label{fig:hybrids}
 \end{figure*}

In this section we investigate the phase evolution of the time-domain inspiral and NR waveforms. Construction of hybrid waveforms, as described in Sec.~\ref{sec:hyb}, demands that the two input waveforms be sufficiently close up to the frequency where they are stitched together.
Following Hannam \textit{et al.}~\cite{Hannam:2007wf,Hannam:2010ec}, we quantify this by aligning the two waveforms in phase at a certain frequency and calculating the phase difference accumulated over a number of GW cycles before this frequency. We choose to make the alignment~\footnote{In Ref.~\cite{Hannam:2007wf} 
the choice made was $M\omega^{}_{\rm GW}=0.1$. We choose a slightly lower frequency so that the \texttt{SEOBv2} waveform without NR calibrations can be generated with sufficient accuracy.}
at GW frequency $M\omega^{}_{\rm GW}=0.075$ or, equivalently, $Mf^{}_{\rm GW}=F\approx 0.012$. The method is the following:
\begin{enumerate}
    \item Given the inspiral and the NR waveforms, we calculate the times where the GW frequency reaches $F=0.012$. We label them as $t=0$ for both the waveforms.
    \item We extract the time-domain phase of the waveforms and make the phases coincide at $t=0$.
    \item We calculate the total phase difference that accumulates over 10 GW cycles in the NR waveform prior to $t=0$, i.e. when the phase (in rad) is $20\pi$ less than that at $t=0$. In other words, if $t=t_0$ is taken to be 10 GW cycles before $t=0$ in the NR waveform, then $\Delta\phi$ is defined as $\phi^{}_{\rm INS}(t_0)-\phi^{}_{\rm NR}(t_0)$.
\end{enumerate}
In Fig.~\ref{fig:phasediff} we show the accumulated phase difference $\Delta\phi=\phi^{}_{\rm INS}(t_0)-\phi^{}_{\rm NR}(t_0)$ (in rad) between the inspiral waveform \texttt{SEOBv2} and the 20 NR waveforms described in Table~\ref{tab:hyb} that are used for the construction of hybrids. We show the results for both \texttt{SEOBv2} and \texttt{SEOBv2\_Horizon}. 
Figure~\ref{fig:phasediff1a} reports the variation with mass ratio, where we highlight the nonspinning cases. In Fig.~\ref{fig:phasediff1b} we show the phase difference as a function of the effective spin parameter $\chi^{}_{\rm eff}=(q\chi^{}_1+\chi^{}_2)/(1+q)$, and highlight the equal-mass binaries. The accumulated phase difference does not exceed 0.5 radians within these 10 cycles for any of these waveforms, implying that the inspiral model is suitable for hybrid construction. With the TH corrections introduced, we see from Fig~\ref{fig:phasediff1a} that the phase difference is lower in most cases, especially for high mass-ratio binaries. This is further demonstrated in Fig.~\ref{fig:phasediff2a}, where we show the difference between the $|\Delta\phi|$ values before and after imposing the TH corrections. To investigate this with a larger set of NR data, we take 218 aligned-spin waveforms from the SXS catalog and repeat the procedure and plot them in Fig.~\ref{fig:phasediff2b} against the mass ratio $q$, and in Fig.~\ref{fig:phasediff2c} against $\chi^{}_{\rm eff}$. Tidal heating introduces only small corrections in the phase; consequently, for symmetric masses and low-spin binaries, the phase error fluctuates. However, towards higher mass ratios (Figure~\ref{fig:phasediff2b}) and high spins (Figure~\ref{fig:phasediff2c}) a statistical behaviour emerges and, for the majority of the binaries, TH corrections contribute to a lesser phase disagreement between the inspiral and the NR waveforms. 

The key takeaway from this section is that the \texttt{SEOBv2\_Horizon} waveform is sufficiently robust for hybrid construction, with phase errors less than 0.5 radians over 10 GW cycles prior to $M\omega^{}_{\rm GW}=0.075$. Furthermore, corrections due to tidal heating of BHs provide a statistical improvement in the phase error for asymmetric masses and highly spinning binaries, especially for anti-aligned spins.

\section{Hybrid construction}\label{sec:hyb}

For constructing a hybrid, the NR waveform has to be `stitched' with the analytical inspiral waveform in a frequency region where both the waveforms have sufficient accuracy, 
which also demands that the morphology of these waveforms do not deviate significantly from one another. For PN waveforms, one conventionally chooses the innermost circular orbit (ISCO) of the binary to be the endmost point of validity. The corresponding GW frequency $f^{}_{\rm ISCO}$ is considered to be the highest frequency for PN expansions. In our work, we ensured that the frequency at the stitching region is below this threshold for all the hybrids.

While stitching at a frequency as low as possible (considering ISCO) should enable one to incorporate the maximum number of NR cycles into the hybrid, there are two more factors that affect the choice of the stitching region:

\begin{itemize}
    \item The initial NR data do not perfectly describe two black holes in quasi-equilibrium. At the start of each simulation, the geometry relaxes to equilibrium on the dynamical time-scale of the individual BHs, changing the mass and spin of each BH by a fractional amount of order $10^{-5}$, and emitting a spurious pulse of gravitational radiation (often referred to as `junk radiation’). 
    \item A robust hybrid should depend weakly on small changes to the stitching region. A monochromatic signal would be completely degenerate under a shift in the coalescence time $t_c$ and the coalescence phase $\phi^{}_c$; this degeneracy is broken by the increase in frequency with time. The start and end of the stitching interval should reflect enough change in the frequency to break this degeneracy. 
\end{itemize}

The first issue is addressed in our work by choosing the stitching region beyond the specified \textit{relaxation time} for each NR simulation in the SXS catalog, 
the time interval (in units of $M$) by which the junk radiation dies out. To address the second requirement, MacDonald \textit{et al.}~\cite{MacDonald:2011ne} have recommended the stitching interval to satisfy $\delta\omega/\omega^{}_m\gtrsim 0.05$, where $\omega^{}_m$ is the GW frequency at the midpoint of the stitching interval, and $\delta\omega$ is the change in frequency over the interval. 
In our work, we perform the stitching over 4 GW cycles, and we place the stitching region in a way to satisfy this condition. In the literature, 
the construction of hybrids has been performed both in time ~\cite{Ajith:2007kx,Husa:2015iqa,Pratten:2020fqn} and  frequency domains~\cite{Santamaria:2010yb,Damour:2010zb}. In our work, we choose the former, primarily because the NR waveforms are available in the time domain, and typically they have too few GW cycles for a reliable Fourier transform.

 In time domain, any two non-precessing binary waveforms $h^A(\Theta,t)$ and $h^B(\Theta,t)$, with the same set of intrinsic parameters $\Theta$, only differ by a relative time shift ($\Delta t$) and an overall constant phase difference ($\phi^{}_0$):
    
\begin{equation*}
    h^A(\Theta,t) = e^{i\phi^{}_0}h^B(\Theta,t+\Delta t).
\end{equation*}
In the region where we expect the PN and NR waveforms to be morphologically similar, we can write

 \begin{equation}
 \label{pnnr}
    h^{\rm PN}(\Theta,t) = e^{i\phi_0}h^{\rm NR}(\Theta,t+\Delta t).
\end{equation}

However, in reality, $h^{\rm PN}(\Theta,t)$ and $h^{\rm NR}(\Theta,t)$ differ from each other morphologically due to waveform systematics, so Eq.~\eqref{pnnr} does not hold in general~\cite{MacDonald:2011ne}. To generate a hybrid waveform, then, one needs to find a suitable region where PN and NR waveforms are approximately the same and minimize the square of the difference between the left-hand and right-hand sides in Eq.~\eqref{pnnr}, by varying $\Delta t$ and $\phi^{}_0$. Since the GW frequency of a CBC increases with time, one can alternatively write the angular GW frequency as a function of time ($\omega(t)$) and minimize the quantity
\begin{equation}
\label{delta}
    \delta=\int_{t_1}^{t_2}\left(\omega^{(\rm NR)}(t)-\omega^{(\rm PN)}(t+\Delta t)\right)^2 \dd t
\end{equation}
with respect to $\Delta t$. Here $t_1$ and $t_2$ denote the start and end of the stitching region.

We list the steps taken to construct the hybrids as follows:

\begin{enumerate}
        
    \item NR data from the SXS catalog sample the GW strain $h^{(\rm NR)}(t)$ in non-uniform timesteps, with a higher sampling rate in regions of higher GW frequency. We resample the data with the coarsest sampling rate by using cubic interpolation. Since our modeling only involves the inspiral parts of the waveforms, this choice does not lead to any loss of information.
    
    \item \texttt{SEOBv2\_Horizon} is generated with the same set of intrinsic parameters that forms the inspiral part of the hybrids, with the same timesteps.
    
    \item Having obtained $h^{(\rm PN)}(t)$ and $h^{(\rm NR)}(t)$, we minimize the quantity $\delta$ defined in Eq.~\eqref{delta} with respect to $\Delta t$. Phase alignments are done at the temporal midpoint of the stitching region, $t=(t_1+t_2)/2$.
    
    \item The hybrid waveform is then constructed as

    \begin{equation}
    \begin{aligned}
          h^{(\rm Hyb)}(t) =&~ \mathcal{F}(t)h^{(\rm PN)}(t)\\ & +\qty[1-\mathcal{F}(t)]e^{i\phi'_0}h^{(\rm NR)}(t+\Delta t'),
    \end{aligned}
    \end{equation}
    where $\phi'_0$ and $\Delta t'$ are the phase offset at the midpoint of the stitching region and the value of $\Delta t$ that minimizes $\delta$, respectively.  $\mathcal{F}(t)$ is a blending function defined by
    
    \begin{equation}
        \begin{aligned}
            \mathcal{F}(t) =
        \left\{
    	    \begin{array}{ll}
    		    1 , &  t < t_1 \\
    		    \frac{1}{2}\qty(1+\cos\frac{\pi(t-t_1)}{t_2-t_1}), & t_1 \leqslant t \leqslant t_2 \\
    		    0 , & t_2 < t\,.
    	    \end{array}
        \right.
        \end{aligned}
    \end{equation}
\end{enumerate}

In the time domain hybrids, parts of the waveforms at $t<t_1$ are purely post-Newtonain, and for $t>t_2$ the waveforms contain purely NR data. In between, the blending function $\mathcal{F}(t)$ smoothly stitches the two waveforms together after proper alignment in phase and time. 

Figure~\ref{fig:hybrids} shows four of the total twenty hybrids made for calibrating the waveform model, at the extreme points of the calibration parameter space. The blue shaded regions correspond to the stitching intervals, $t_1\leqslant t \leqslant t_2$. In Table~\ref{tab:hyb}, we list the NR data used to create the set of hybrids for calibration. We report the simulation labels of aligned-spin NR data from SXS, dimensionless spin magnitudes for the two component BHs, the effective spin parameter used for modeling the inspiral (discussed in Sec.~\ref{sec:correspondence}), the number of orbits present in the NR data, and the value of dimensionless frequency $Mf$ at the midpoint of the stitching regions.

\section{Modeling the hybrids in the frequency domain}\label{sec:model}

We perform phenomenological parametrization of the hybrid waveforms in the frequency domain. Template waveforms in the frequency domain are of particular preference since the parametrization is easier, and also a search using frequency-domain templates is computationally inexpensive compared to one using time-domain templates. Our parametrization follows the construction of \texttt{PhenomD}~\cite{Husa:2015iqa,Khan:2015jqa} for modeling the phase and amplitude of the frequency-domain hybrids. 

\texttt{PhenomD} used uncalibrated \texttt{SEOBNRv2} as the inspiral approximant, while 19 NR waveforms extracted by SpEC and BAM~\cite{Bruegmann:2006ulg} codes were used for hybrid construction and calibration. The model was constructed in the frequency domain by dividing the entire waveform into three parts -- inspiral, intermediate, and merger-ringdown. 
This modular nature allows one to use a different inspiral model while keeping the merger and ringdown intact. In that model, inspiral is defined to be the region below  $F=0.018$. Merger and ringdown phases are separated by identifying the ringdown frequencies of different hybrids that are used to calibrate the model. In \texttt{PhenomD}, the end of inspiral ($F=0.018$) was chosen in such a way that beyond that frequency, each of the hybrids contained purely NR data. The merger-ringdown model of \texttt{PhenomD}, then, is a model of purely NR data, which we do not need to recalibrate in this work. In fact, as discussed in Ref.~\cite{Husa:2015iqa}, the modular nature of \texttt{PhenomD} allows independent modeling of the inspiral, intermediate, and merger-ringdown parts separately, without even needing any hybrid waveform. The hybrid dataset, however, serves as a benchmark to compare the final model with and provide long time-domain waveforms for better frequency-domain modeling. Additionally, the hybrids are created with a smooth transition of phase and amplitude by the stitching process, producing accurate values of the pseudo-PN parameters capturing higher-order terms in frequency.

\subsection{Inspiral phase model}

\begin{figure*}
\centering
 \subfigure[]{\includegraphics[width=0.245\textwidth]{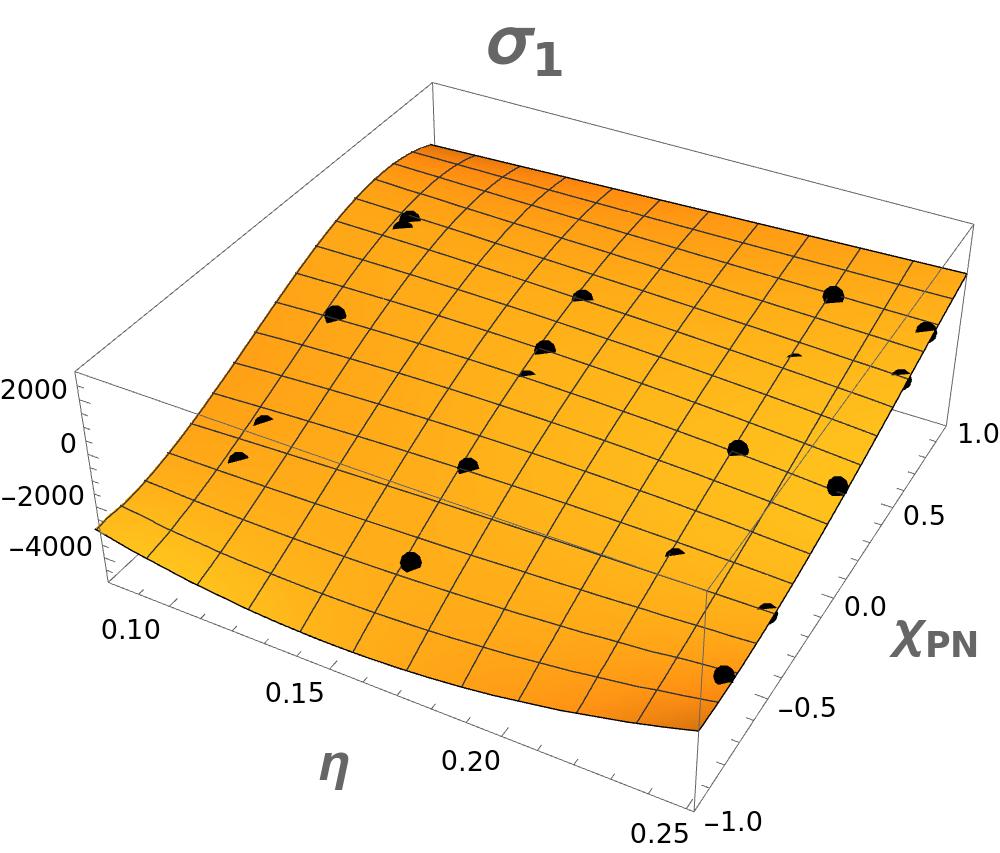}}
 \hfill
 \subfigure[]{\includegraphics[width=0.245\textwidth]{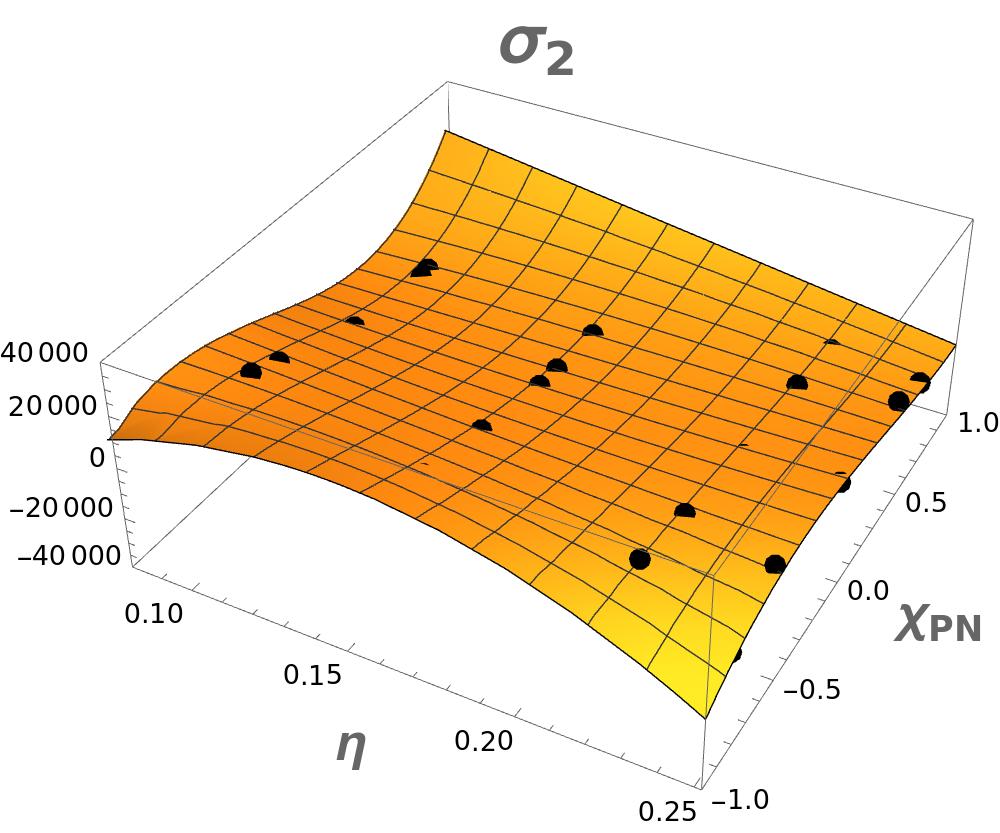}}
 \hfill
 \subfigure[]{\includegraphics[width=0.245\textwidth]{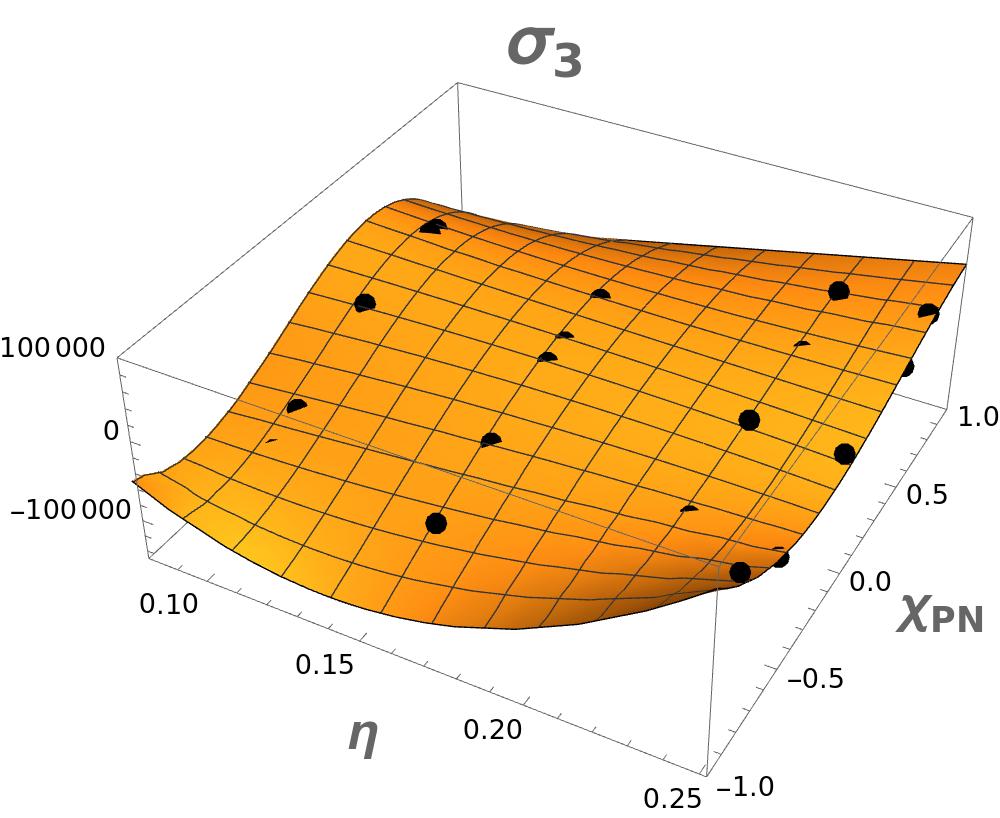}}
 \hfill
 \subfigure[]{\includegraphics[width=0.245\textwidth]{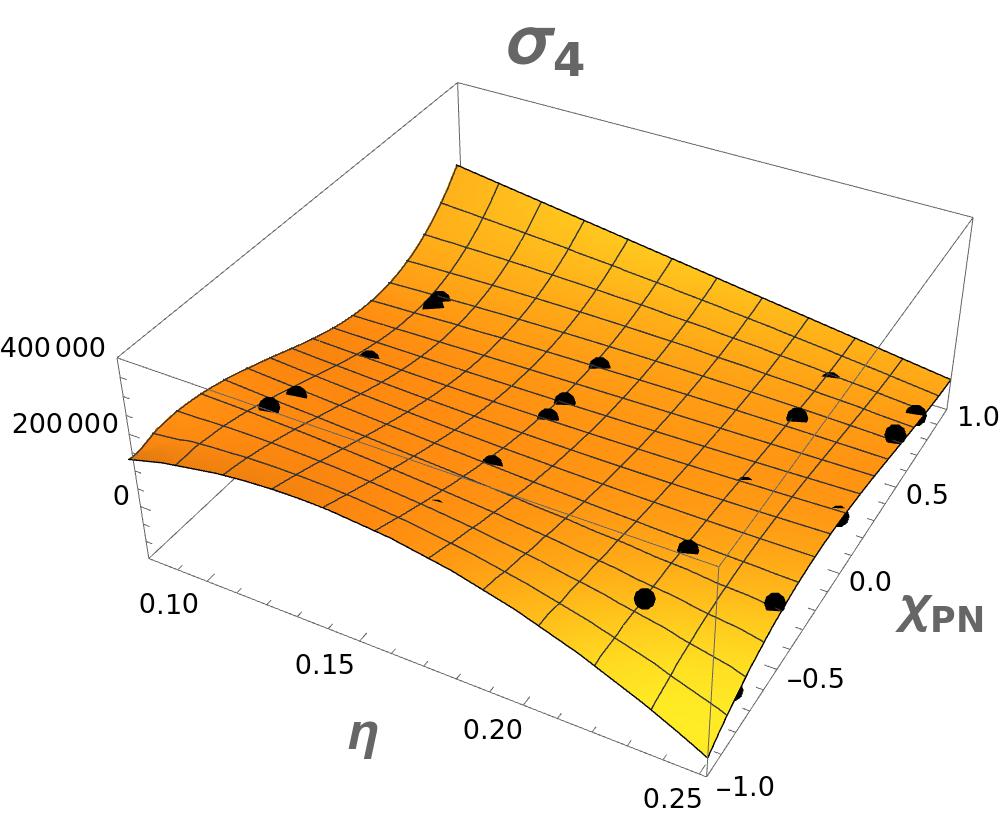}}
 \hfill
 \caption{Fits of $\sigma^{}_i$ calculated by the \texttt{NonLinearModelFit} module of Mathematica. The surfaces correspond to the 2D fits with $\eta$ and $\chi^{}_{\rm PN}$ as described in Eq.~\eqref{eq:Lambda}.}
\label{fig:sigma}
 \end{figure*}
 
The inspiral approximant used in the hybrid data is \texttt{SEOBv2\_Horizon}, as discussed earlier. Since the final model is a phenomenological one, it requires ready-to-use analytical forms for the phase and amplitude with extra coefficients calibrated to the hybrid data.  
We write the frequency-domain phase of a waveform in the inspiral as 
\begin{equation}\label{eq:phiins}
\begin{aligned}
     \Psi^{}_{\rm INS}(F;\Theta) =&~ \Psi^{}_{\rm{TF2}}(F;\Theta) + \Psi^{}_{\rm TH}(F;\Theta) \\&+ \frac{1}{\eta}\left[\sigma^{}_0 + \sigma^{}_1 F + \frac{3}{4}\sigma^{}_2 F^{4/3}\right. \\& \left. \quad\quad + \frac{3}{5}\sigma^{}_3 F^{5/3} + \frac{1}{2}\sigma^{}_4 F^2 \right],
\end{aligned}
\end{equation}
where $\Psi^{}_{\rm{TF2}}(F;\Theta)$ and $\Psi^{}_{\rm TH}(F;\Theta)$ are 3.5PN phase of \texttt{TaylorF2} (described in Appendix~\ref{app:tf2phase}) and the TH correction (Eq.~\eqref{eq:phase correction}), respectively. $\sigma^{}_i\>\> (i=0-4)$ are 5 phenomenological pseudo-PN parameters. The analytical ansatz in Eq.~\eqref{eq:phiins} is the same as in \texttt{PhenomD} inspiral model. 
Since we want to leverage the modular nature of \texttt{PhenomD} to concatenate its merger-ringdown model with our own inspiral model, the end of the inspiral phase in the current work needs to remain unchanged, which is $F=0.018$. We also note here that one can, in principle, directly add the correction $\Psi^{}_{\rm TH}(F;\Theta)$ to the \texttt{PhenomD} inspiral phase (as is done for the amplitude, discussed in Sec.~\ref{sec:amplitude}), without any recalibration. However, since we intend to avoid adding the correction to the NR information present in the last few cycles of inspiral, recalibrating the phase allows a smooth transition from PN to NR. To find the parameter values corresponding to each hybrid waveform, we fit the first derivative of the phase with respect to $F$:
\begin{equation}\label{eq:phiprimeins}
\begin{aligned}
     \Psi^\prime_{\rm INS}(F;\Theta) =&~ \Psi^\prime_{\rm{TF2}}(F;\Theta) + \Psi^\prime_{\rm TH}(F;\Theta) \\&+ \frac{1}{\eta}\left[\sigma^{}_1 + \sigma^{}_2 F^{1/3}\right. \\& \left. \quad\quad + \sigma^{}_3 F^{2/3} + \sigma^{}_4 F \right],
\end{aligned}
\end{equation}
where $\Psi^\prime= \partial\Psi/\partial F$. $\sigma_0$ is determined by imposing $C^{(1)}$ continuity in phase at the boundary between inspiral and merger. The fits are performed over the frequency range $F\in [0.0035,0.019]$, ending at a slightly higher frequency, to reduce boundary effects at the interface and find robust fits for $\sigma_i$. For all the 20 hybrids used for calibration, the stitching regions are placed within this frequency range, which warrants that there is purely NR data beyond $F=0.018$.

\subsection{Inspiral and intermediate amplitude model}\label{sec:amplitude}

To model the amplitude within the frequency range of the inspiral, we build the inspiral amplitude model by adding the correction due to TH, given by Eq.~\eqref{eq:amp correction}, to the \texttt{PhenomD} amplitude: 
\begin{equation}
\label{eq:ampmodel}
    \Tilde{A}^{}_{\rm INS}(F) = \Tilde{A}^{}_{\rm D}(F)+\Tilde{A}^{}_{\rm TH}(F)\,.
\end{equation}
We do not recalibrate the amplitude pseudo-PN parameters of \texttt{PhenomD}, since the inspiral amplitude model ends (at $F=0.014$) before the phase model does (at $F=0.018$), and the amount of NR information is less than that of phase. Moreover, we find that the effect of the amplitude correction is significantly less than the phase correction, diminishing the necessity of a recalibration. 

At the interface of two frequency intervals modeled separately, $C^{(1)}$ continuity on the amplitude cannot be imposed in a straightforward manner. For the frequency-domain phase, one has the freedom to tune the coalescence time $t^{}_c$ and the coalescence phase $\phi^{}_c$ to impose continuity of the phase and its derivative across an interface. For amplitude, however, there is no such freedom. If the amplitude is modeled separately in different frequency regimes, a different strategy has to be followed. To ensure that the amplitude and its derivative are continuous throughout the three phases, one needs a frequency interval to fit a polynomial that satisfies these conditions at the beginning and the end. This `intermediate' interval is defined to be $F\in [0.014,F^{}_{\rm peak}]$, where $F^{}_{\rm peak}\equiv Mf^{}_{\rm peak}$ is the frequency corresponding to the peak amplitude. For $F<0.014$, the amplitude is given by Eq.~\eqref{eq:ampmodel}. In the intermediate region, the amplitude is approximated as a polynomial in $f$:

\begin{equation}
\begin{aligned}
\label{eq:ampint}
     \Tilde{A}^{}_{\rm int} = A_0 & \left[\delta^{}_0 + \delta^{}_1 f + \delta^{}_2 f^2 \right. \\ 
                     & \left. +\delta^{}_3 f^3 + \delta^{}_4 f^4\right]. 
\end{aligned}
\end{equation}
Here $A^{}_0$ includes the leading order $f^{-7/6}$ behaviour. Evaluation of $\delta^{}_i$ follows the steps of \texttt{PhenomD}, described in Sec. V(C) of Ref~\cite{Khan:2015jqa}. We briefly summarize it here:
\begin{itemize}
    \item Equation~\eqref{eq:ampint} has five parameters, requiring five independent equations for unique solutions. Two of them come from the $C^{(1)}$ continuity of $\Tilde{A}_{\rm int}$ with $\Tilde{A}_{\rm INS}(f)$ at the beginning ($F=0.014$), and two from $C^{(1)}$ continuity with \texttt{PhenomD} amplitude at $F^{}_{\rm peak}$. The fifth equation appears from an additional condition that the polynomial coincides with \texttt{PhenomD} amplitude at the mid-frequency (chosen as a collocation point), $F=(0.014+F_{\rm peak})/2$.
    \item Solving the aforementioned set of equations, one finds the form of $\Tilde{A}^{}_{\rm int}$, which smoothly connects the intermediate region with inspiral and merger.
    \item It is worth mentioning that while choosing more collocation points would improve the model accuracy, it would also encumber the model with a larger set of equations to solve. Keeping in mind that amplitude errors are less consequential than phase errors, we refrain from considering more collocation points.
\end{itemize}

In Appendix~\ref{app:ringdown}, we describe the evaluation of the peak frequency $f^{}_{\rm peak}$ and the ringdown frequency $f^{}_{\rm RD}$ as functions of the BH masses and spins.

\subsection{Correspondence between the phenomenological and physical parameters}\label{sec:correspondence}

\begin{figure*}
\centering
  \includegraphics[width=\textwidth]{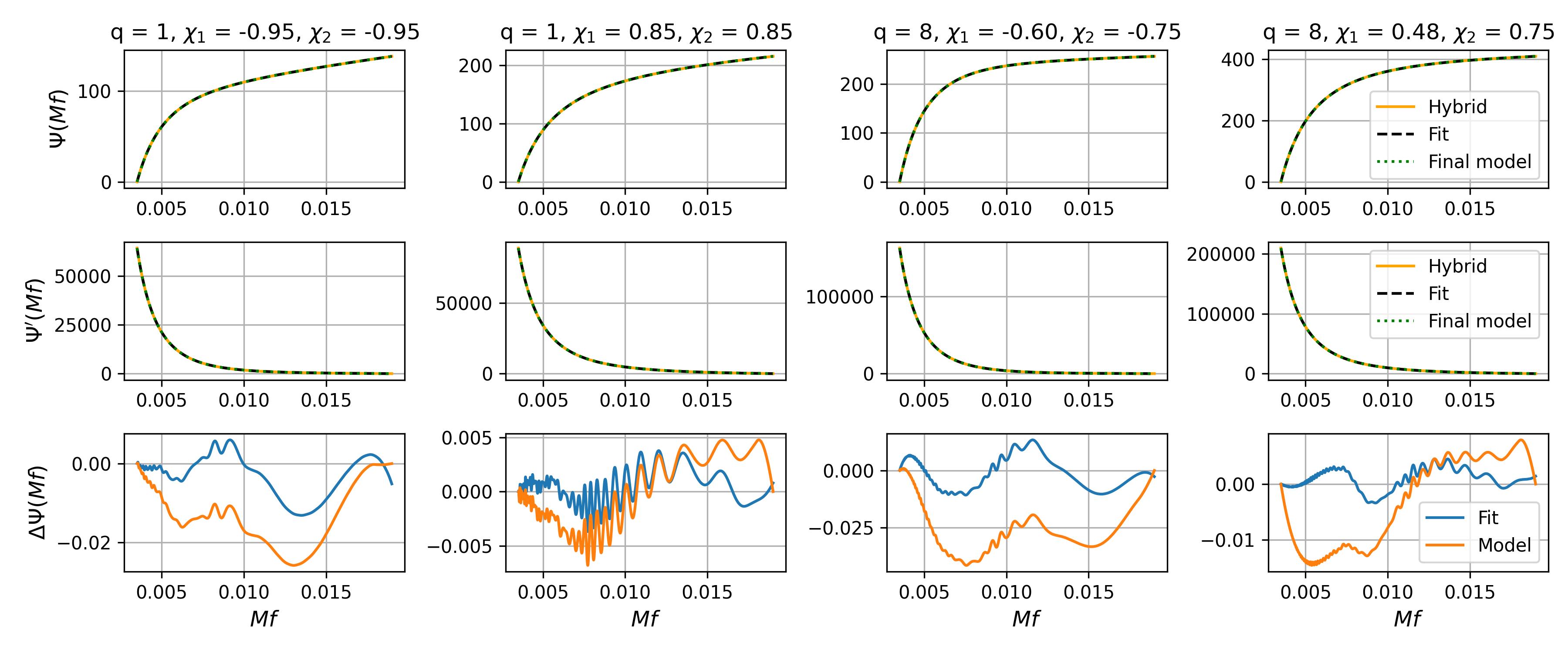}
  \caption{Phase, phase derivative, and phase difference between the hybrids (same as in Fig.~\ref{fig:hybrids}), individual fits obtained from Eq.~\eqref{eq:phiprimeins}, and the final model prediction. \textit{Top panels}: Inspiral phase $\Psi (Mf)$. \textit{Middle panels}: $\Psi^\prime (Mf)=\partial\Psi/\partial (Mf)$ plotted against $Mf$.  \textit{Bottom panels}: $\Delta \Psi^{}_{\rm Fit} = (\Psi^{}_{\rm Hybrid} - \Psi^{}_{\rm Fit})$, and $\Delta \Psi^{}_{\rm Model} = (\Psi^{}_{\rm Hybrid} - \Psi^{}_{\rm Model})$. All the plots are shown within the inspiral frequency range, $0.0035\leqslant Mf \leqslant 0.018$.}
\label{fig:phase}
 \end{figure*}

\begin{figure}
\centering
 \subfigure[]{\includegraphics[width=0.493\linewidth]{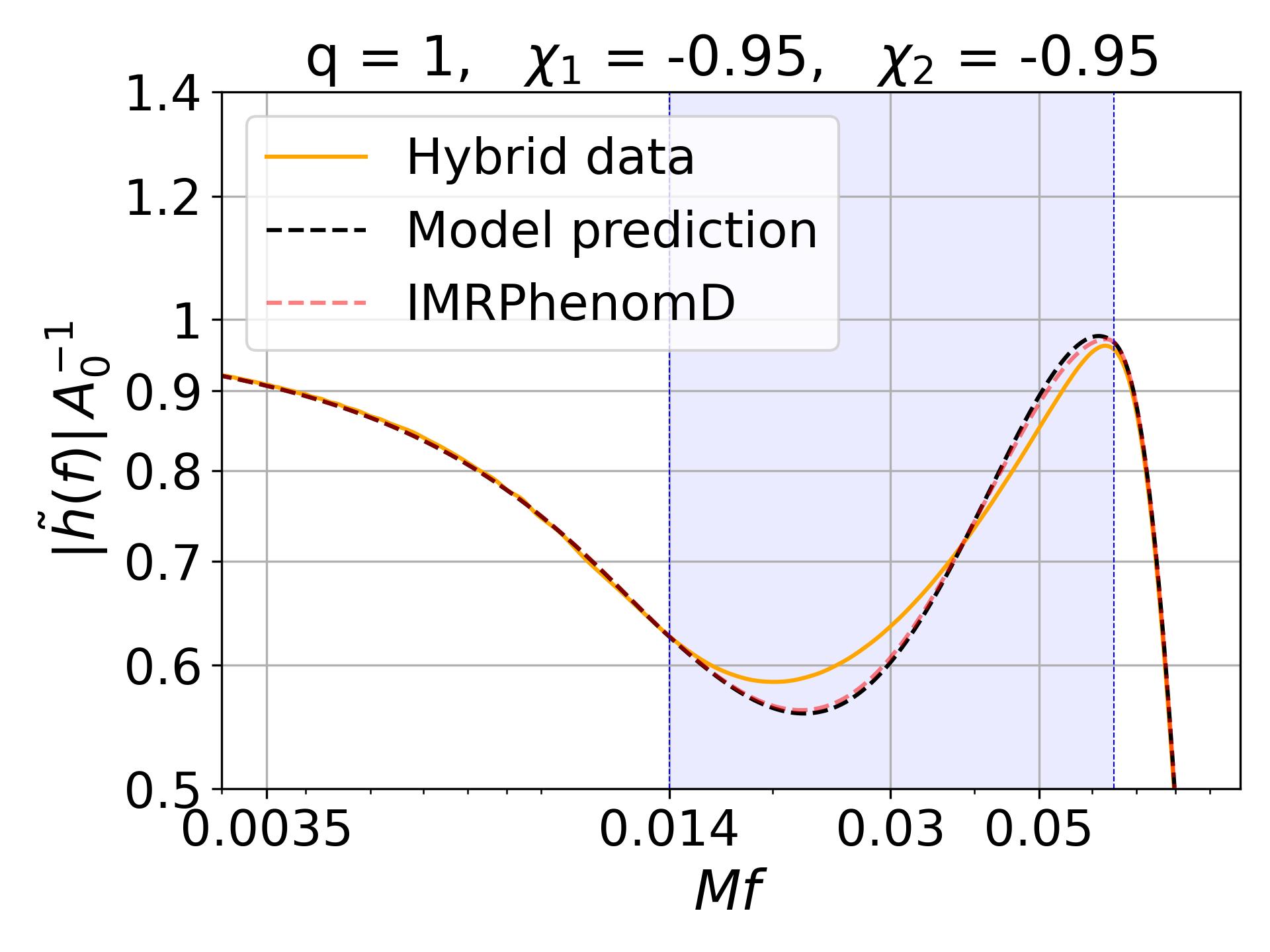}}
 \hfill
 \subfigure[]{\includegraphics[width=0.493\linewidth]{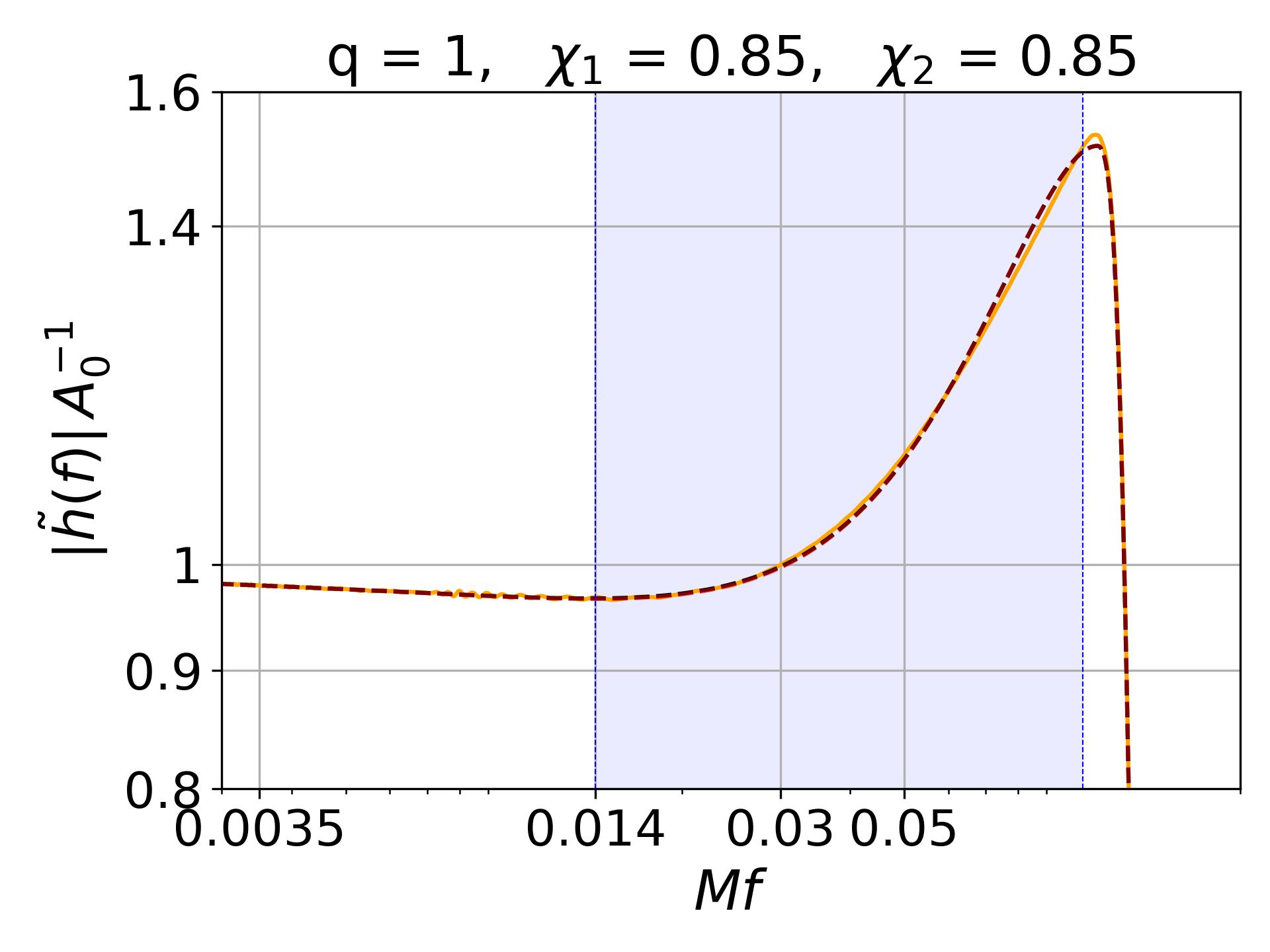}}
 \hfill
 \subfigure[]{\includegraphics[width=0.493\linewidth]{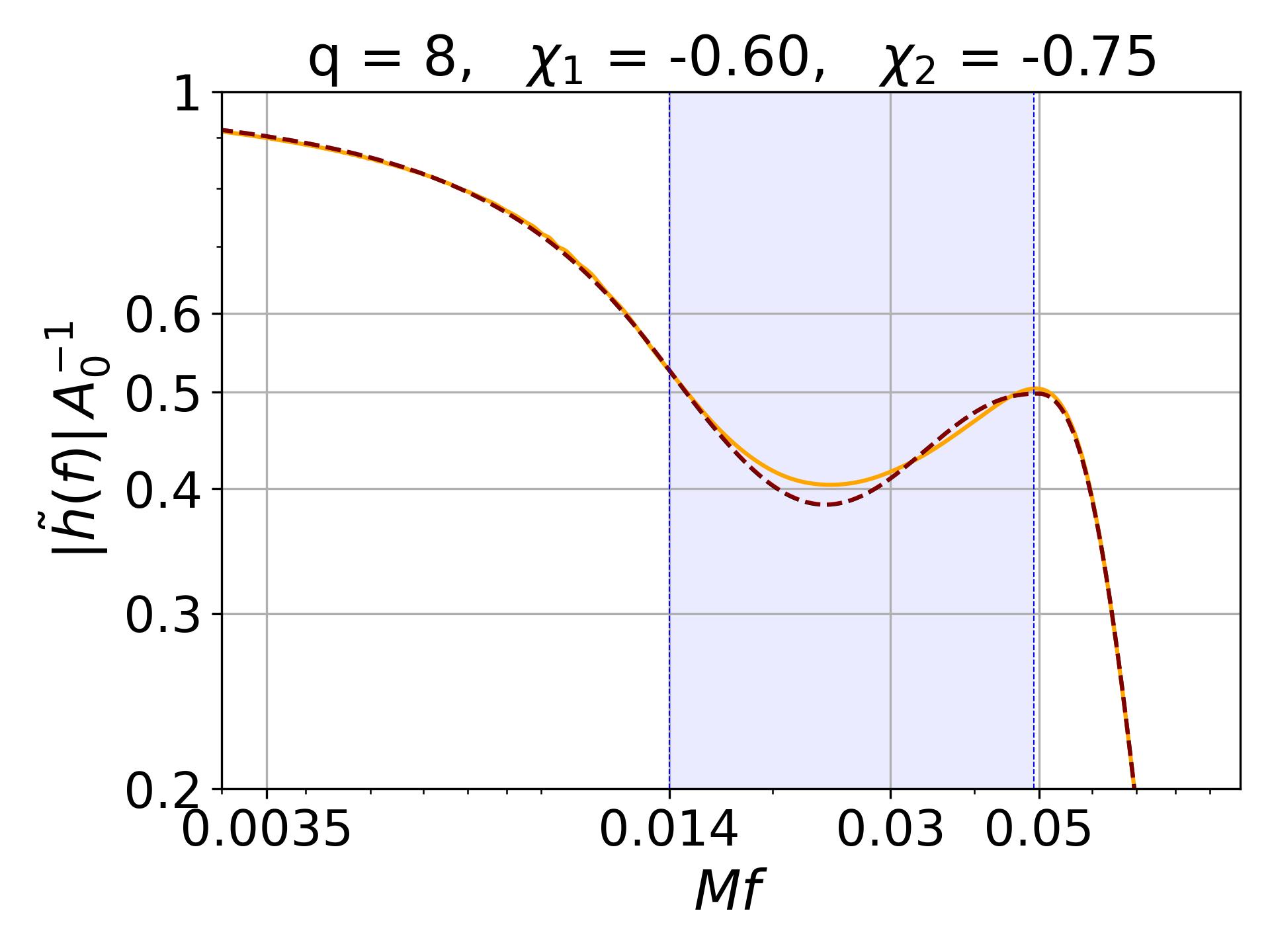}}
 \hfill
 \subfigure[]{\includegraphics[width=0.493\linewidth]{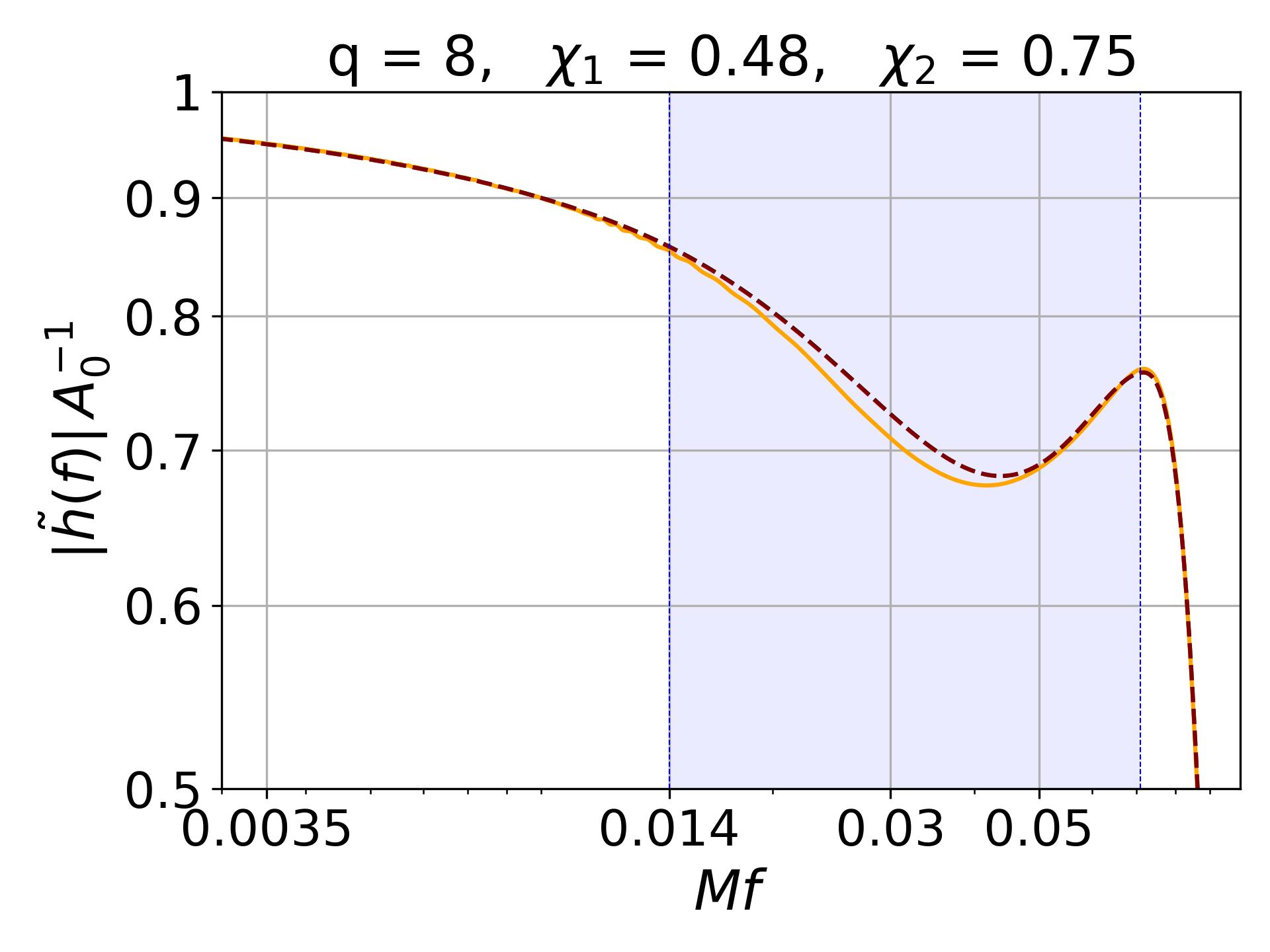}}
 \caption{Rescaled amplitude of the full IMR waveforms as functions of $Mf$. The blue shaded regions denote the intermediate frequency range, $0.014\leqslant Mf \leqslant F^{}_{\rm peak}$.}
\label{fig:imramp}
\end{figure}

Our model has 4 phenomenological parameters corresponding to the inspiral phase. These parameters play an intermediate role in generating gravitational waveforms of aligned-spin compact binaries. To generate waveforms for arbitrary values of the masses and spins, one needs to build a correspondence between the set of physical parameters $\{\eta,\chi^{}_1,\chi^{}_2\}$ and the set of phenomenological parameters $\{\sigma^{}_i\}$. The total mass $M$ of the binary works as a trivial scaling factor.

In PN expansions, the leading order contribution of spin in the phase of the waveform appears as a function of the combination
\begin{equation}
\label{eq:eff_spin}
    \chi^{}_{\rm PN} = \chi^{}_{\rm eff} - \frac{38\eta}{113}(\chi^{}_1+\chi^{}_2),
\end{equation}
where
\begin{equation}
    \chi^{}_{\rm eff}=\frac{m^{}_1\chi^{}_1+m^{}_2\chi^{}_2}{M}.
\end{equation}

In this model, $\chi^{}_{\rm PN}$ is used as a single spin parameter to generate the phenomenological parameters. This ``effective spin approximation" works well for aligned-spin binaries, especially in the inspiral regime~\cite{Khan:2015jqa}.

To establish the correspondence between physical and phenomenological parameters, we fit $\sigma^{}_i$ obtained from different hybrids with a polynomial of $\eta$ (up to second order) and the effective spin parameter $\chi^{}_{\rm PN}$ (up to third order):
\begin{equation}\label{eq:Lambda}
\begin{aligned}   
    \sigma^{}_i = ~ \lambda_{00}^i+\lambda_{10}^i\eta ~& +(\chi^{}_{\rm PN}-1)(\lambda_{01}^i+\lambda_{11}^i\eta+\lambda_{21}^i\eta^2)\\
                & +(\chi^{}_{\rm PN}-1)^2(\lambda_{02}^i+\lambda_{12}^i\eta+\lambda_{22}^i\eta^2)\\
                & +(\chi^{}_{\rm PN}-1)^3(\lambda_{03}^i+\lambda_{13}^i\eta+\lambda_{23}^i\eta^2).
\end{aligned}
\end{equation}
For each $\sigma^{}_i$, this polynomial generates a set $\{\lambda^i_{jk}\}$ to build the correspondence between physical and phenomenological parameter sets. Figure~\ref{fig:sigma} shows the surfaces of $\sigma^{}_i$ as functions of $\{\eta,\chi^{}_{\rm PN}\}$ with the ansatz in Eq.~\eqref{eq:Lambda}. These best-fit surfaces generate a constant set of $\{\lambda^i_{jk}\}$ which define the inspiral model, along with the \texttt{TaylorF2} phase and the TH correction. In Appendix~\ref{app:lambda}, we list the coefficient values corresponding to each $\sigma^{}_i$.
A schematic flowchart for generating the inspiral phase model starting from the intrinsic binary parameters can be described as:
\begin{equation*}
    {\{\eta,\chi^{}_1,\chi^{}_2\}}\xrightarrow[\text{spin}]{\text{Effective}}\{\eta,\chi^{}_{\rm PN}\}\xrightarrow{\lambda^i}{\{\sigma^{}_i\}}\xrightarrow{}{\text{Inspiral model}}.
\end{equation*}

\subsection{Full IMR model}\label{sec:imr}

For the phase, we have separately modeled the frequency region $0.0035\leqslant F \leqslant 0.018$, and we have the phase model of \texttt{PhenomD} thereafter. For the amplitude, we have a corrected model in the range $F \leqslant 0.014$, and a polynomial function for $0.014\leqslant F \leqslant F^{}_{\rm peak}$. This gives us one interface for the phase and two interfaces for the amplitude to impose $C^{(1)}$ continuity. For the amplitude, however, $C^{(1)}$ continuity is ensured by the intermediate polynomial itself. For the phase, we can vary $t_c$ and $\phi_c$ of the inspiral phase. After imposing $C^{(1)}$ continuity, we can generate the complete IMR phase and amplitude by defining a step function

\begin{equation}
        \begin{aligned}
            \theta (f-f_0) =
        \left\{
    	    \begin{array}{ll}
    		    -1 , &  f < f^{}_0 \\
    		    +1 , & f\geqslant f^{}_0\,.
    	    \end{array}
        \right.
        \end{aligned}
\end{equation}
Using this function, one can define
\begin{equation}
    \theta ^\pm (f;f^{}_0)=\frac{1}{2}\qty[1\pm \theta (f-f^{}_0)],
\end{equation}
so that the IMR phase can be written as
\begin{equation}\label{eq:phiimr1}
    \Phi^{}_{\rm IMR}(F)=\Phi^{}_{\rm INS}(F)\theta^-(F;F^{}_2) + \theta^+(F;F^{}_2)\Phi^{}_{\rm D}(F),
\end{equation}
where $F^{}_2=0.018$, and $\Phi^{}_{\rm D}(F)$ corresponds to the phase model of \texttt{PhenomD}. $\Phi^{}_{\rm INS}(F)$ is given by Eq.~\eqref{eq:phiins}.

The IMR amplitude model follows a similar treatment, which can be expressed as
\begin{equation}\label{eq:ampimr}
\begin{aligned}
    \Tilde{A}^{}_{\rm IMR}(F) & = \Tilde{A}^{}_{\rm INS}(F)\theta^-(F;F^{}_3) \\ + & \theta^+(F;F^{}_3)\Tilde{A}^{}_{\rm int}(F)\theta^-(F;F^{}_4) + \theta^+(F;F^{}_4)\Tilde{A}^{}_{\rm D}(F),
\end{aligned}
\end{equation}
where $F^{}_3=0.014$, and $F^{}_4=F_{\rm peak}$. $\Tilde{A}^{}_{\rm INS}(F)$ is the inspiral amplitude in Eq.~\eqref{eq:ampmodel}, $\Tilde{A}^{}_{\rm int}(F)$ is the intermediate amplitude defined in Eq.~\eqref{eq:ampint}, and $\Tilde{A}^{}_{\rm D}(F)$ is the \texttt{PhenomD} amplitude.

Finally, the plus and cross polarization states of the complete frequency-domain waveform read,
\begin{equation}
\begin{aligned}
    \Tilde{h}_+(f) = & \Tilde{A}^{}_{\rm IMR}(f)\qty(\frac{1+\cos^2\iota}{2}) \\ & \times\exp{-i(\Phi^{}_{\rm IMR}(f)-\phi^{}_0-2\pi ft^{}_0)},
\end{aligned}
\end{equation}
and
\begin{equation}
\begin{aligned}
    \Tilde{h}_{\cross}(f) = & -i\Tilde{A}^{}_{\rm IMR}(f)\cos\iota \\ & \times\exp{-i(\Phi^{}_{\rm IMR}(f)-\phi^{}_0-2\pi ft^{}_0)}.
\end{aligned}
\end{equation}
Here $\iota$ is the angle of inclination of the binary plane to the line of sight, $\phi^{}_0$ and $t^{}_0$ carry the overall phase and timeshift freedom of the complete waveform. We call the final waveform model \texttt{IMRPhenomD\_Horizon}, and abbreviate it to \texttt{PhenomD\_Horizon}.

In Fig.~\ref{fig:phase}, we show the inspiral phase $\Psi(Mf)$, its derivative with respect to $Mf$, and the difference between the phase of the hybrid data and the model, for the same configurations as in Fig.~\ref{fig:hybrids}. We show both the individual fits from the ansatz in Eq.~\eqref{eq:phiins} and the final model prediction obtained from the best fit values of ${\lambda^i_{jk}}$ from Eq.~\eqref{eq:Lambda}. In Fig.~\ref{fig:imramp}, we show the inspiral and intermediate amplitude, rescaled by the leading order factor $A^{}_0=\frac{1}{\pi^{2/3}}\sqrt{5\eta/24}f^{-7/6}$. The intermediate regions are shown in a blue shaded colour. We also show the amplitude of \texttt{PhenomD} on the same plots.

\section{Faithfulness of the model}\label{sec:model validation}

\begin{table}
\vspace{4.5mm}

    \centering
    \begin{tabular}{p{0.03\textwidth}p{0.16\textwidth}p{0.08\textwidth}p{0.08\textwidth}p{0.08\textwidth}}
        \hline
        \# & Simulation label & $q$ & $\chi^{}_1$ & $\chi^{}_2$ \\
        \vspace{0.5mm}\\
        \hline
        \hline
        \vspace{0.5mm}\\
        1 & SXS:BBH:0159 & 1 & $-$0.9 & $-$0.9 \\
        2 & SXS:BBH:0154 & 1 & $-$0.8 & $-$0.8 \\
        3 & SXS:BBH:0148 & 1 & $-$0.44 & $-$0.44 \\
        4 & SXS:BBH:0150 & 1 & 0.2 & 0.2 \\
        5 & SXS:BBH:0170 & 1 & 0.44 & 0.44 \\
        6 & SXS:BBH:0155 & 1 & 0.8 & 0.8 \\
        7 & SXS:BBH:0160 & 1 & 0.9 & 0.9 \\
        8 & SXS:BBH:0157 & 1 & 0.95 & 0.95 \\
        9 & SXS:BBH:0014 & 1.5 & $-$0.5 & 0 \\
        10 & SXS:BBH:0008 & 1.5 & 0 & 0 \\
        11 & SXS:BBH:0013 & 1.5 & 0.5 & 0 \\
        12 & SXS:BBH:0046 & 3 & $-$0.5 & $-$0.5 \\
        13 & SXS:BBH:0036 & 3 & $-$0.5 & 0 \\
        14 & SXS:BBH:0168 & 3 & 0 & 0\\
        15 & SXS:BBH:0031 & 3 & 0.5 & 0 \\
        16 & SXS:BBH:0047 & 3 & 0.5 & 0.5 \\
        17 & SXS:BBH:0056 & 5 & 0 & 0 \\
        18 & SXS:BBH:0181 & 6 & 0 & 0 \\
        19 & SXS:BBH:1424 & 6.465 & $-$0.66 & $-$0.8 \\
        20 & SXS:BBH:0298 & 7 & 0 & 0 \\
        \vspace{0.5mm}\\
        \hline
    
    \end{tabular}
    \caption{List of the extra hybrids created for validating the model.}
    \label{tab:hyb_nc}
\end{table}

\begin{figure}[t]
    \centering
    \includegraphics[width=0.98\linewidth]{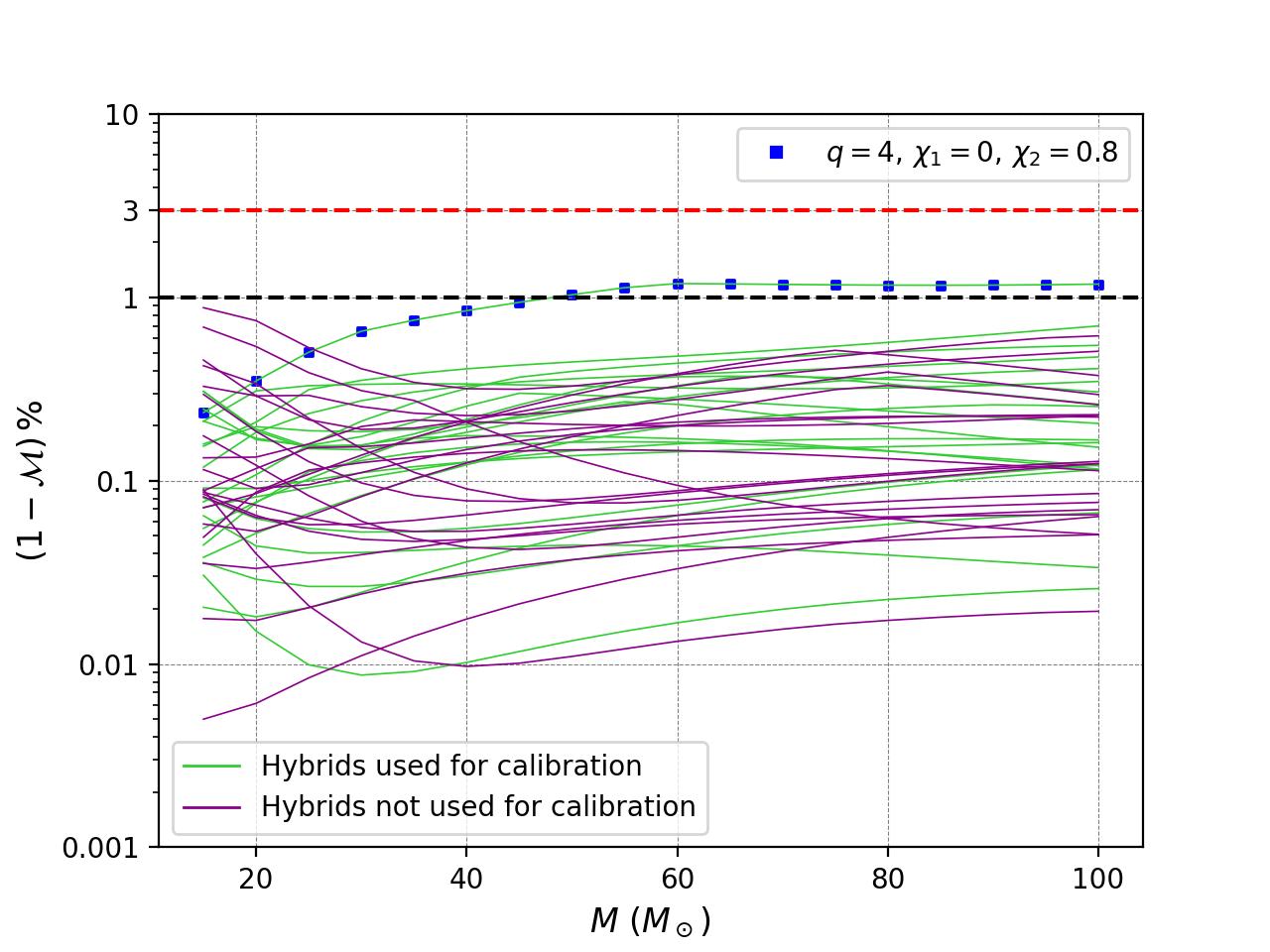}
    \caption{Mismatches (\%) between \texttt{IMRPhenomD\_Horizon} and the hybrid waveforms (Table~\ref{tab:hyb} and Table~\ref{tab:hyb_nc}) in aLIGO ZDHP noise PSD with a lower cutoff of 10 Hz, as a function of the total mass.}
    \label{fig:mismatch}
\end{figure}

A phenomenological model, calibrated with a set of ``target hybrids", should be compared with a larger set of ``test hybrids" to assess its accuracy. The test hybrids should contain both the target hybrids and a new set of hybrids interspersed within the parameter space of calibration. To study the model's behavior beyond the calibration range, test hybrids can be extended further in the parameter space. How well the final model can predict these test hybrids is denoted by its \textit{effectualness} and \textit{faithfulness}~\cite{Ajith:2007kx,Lindblom:2008cm}. A model is effectual if it is accurate enough to predict a GW signal from the detector noise, while to be faithful it also has to have enough accuracy to estimate the binary parameters. A faithful model is also effectual, but the converse may not be true. Given any two signals $h(t)$ and $g(t)$, buried in the noise $n(t)$, one can define a noise-weighted inner product
\begin{equation}
    \braket{h}{g}=2\int_{0}^{\infty}\frac{\Tilde{h}(f)\Tilde{g}^\ast (f)+\Tilde{h}^\ast (f)\Tilde{g}(f)}{S_n(f)}\dd f\,,
\end{equation}
where $\Tilde{h}(f)$ is the Fourier transform of $h(t)$ defined by
\begin{equation}
    \Tilde{h}(f)=\int_{-\infty}^{\infty}h(t)e^{-2\pi i ft}\dd t\,,
\end{equation}
and $S_n(f)$ is the power spectral density (PSD) of the noise.

Equipped with this definition of an inner product between two waveforms, one can define the \textit{match} between them by maximizing the inner product with respect to the constant phase and timeshift freedom between the waveforms $\Tilde{h}(f)$ and $\Tilde{g}(f)$, normalized by their individual norms:
\begin{equation}
   \mathcal{M}=\underset{t_c,\phi_c}{\text{min}}\,\,\frac{\braket{h(\lambda)}{g(\lambda ')}}{\sqrt{\braket{h(\lambda)}{h(\lambda)}\braket{g(\lambda ')}{g(\lambda ')}}}\,.
\end{equation}
Here $\lambda$ and $\lambda '$ denote the intrinsic parameters. The \textit{mismatch}, defined as $1-\mathcal{M}$, quantifies how much the waveform $\Tilde{h}(f)$ morphologically differs from $\Tilde{g}(f)$. The faithfulness of a model is quantified by demanding that the mismatches between the model and the validation dataset (test hybrids) do not exceed a certain threshold. For \texttt{PhenomD}, the threshold was 1\%.

Since the total mass of the binary works as a scaling factor for the gravitational waveforms, it defines how many GW cycles are contained within the sensitive frequency band of a detector. The lower the total mass, the more time the binary spends in its inspiral phase, resulting in a longer waveform given a lower cutoff for the frequency. In Advanced LIGO, complete IMR models become important above a critical value $\sim 12M_\odot$~\cite{Buonanno:2009zt} of the binary mass.
To assess our model's faithfulness, we create a set of test hybrids that are long enough to cover the sensitive frequency band of Advanced LIGO for a total binary mass as low as $10M_\odot$. At 10 Hz, the dimensionless frequency corresponding to a $10M_\odot$ binary is $Mf\approx 0.0005$. 

\begin{figure*}[btp]
\centering
  \subfigure[]{\label{fig:mismatch_TF2_HTF2_LIGO}\includegraphics[width=0.489\textwidth]{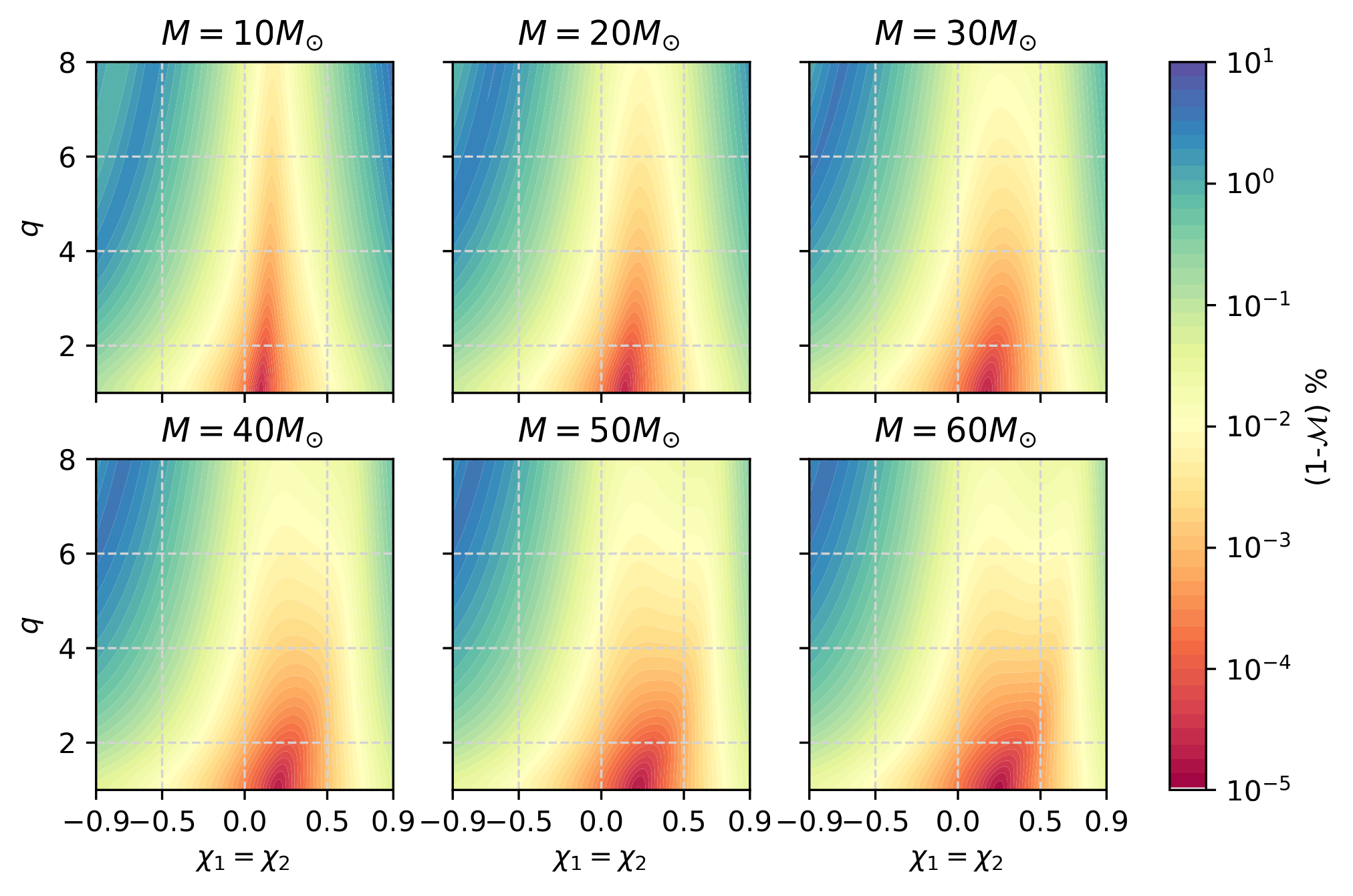}}
 \hfill
 \subfigure[]{\label{fig:mismatch_TF2_HTF2_isco_LIGO}\includegraphics[width=0.489\textwidth]{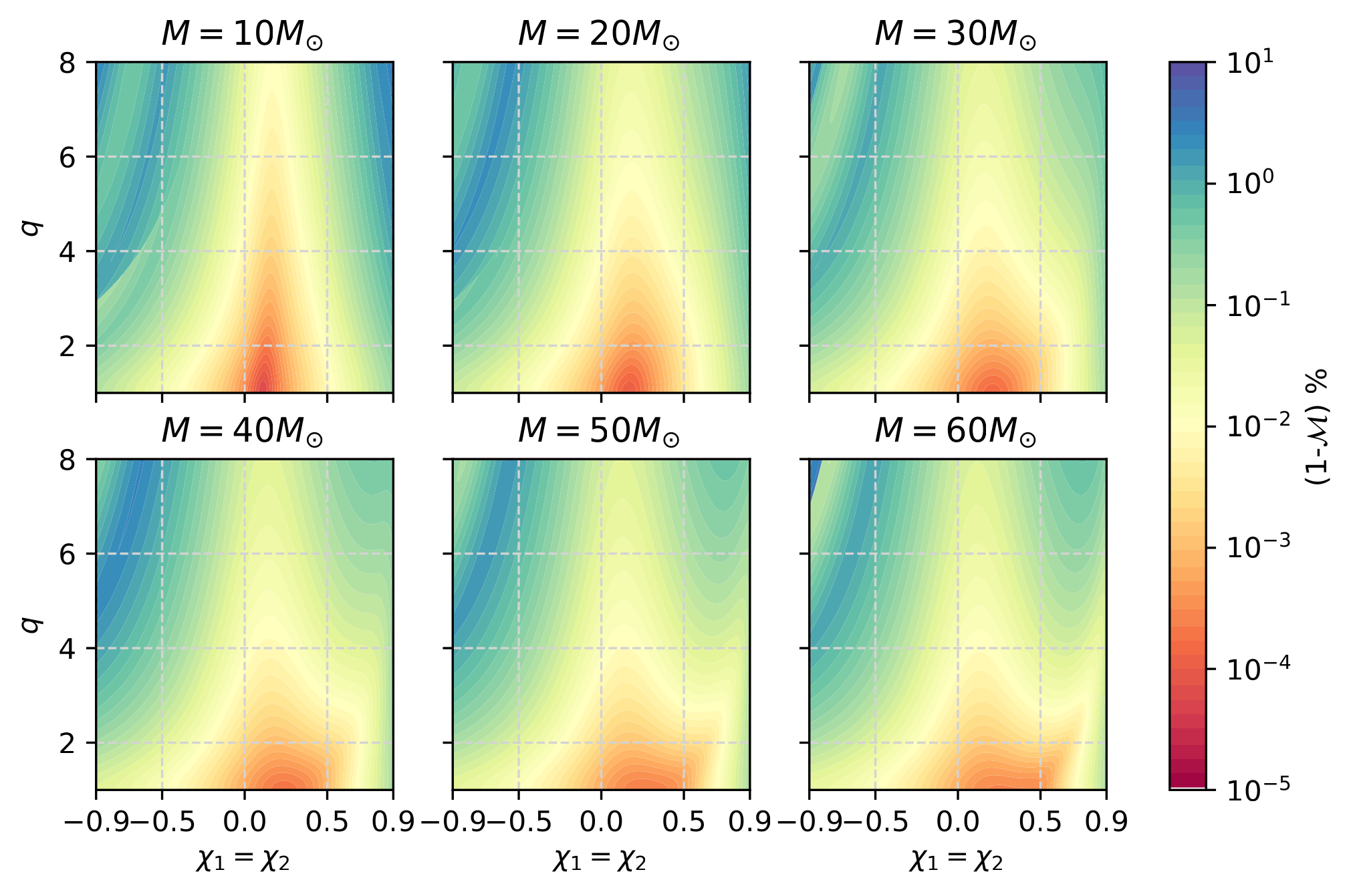}}
 \hfill
 \subfigure[]{\label{fig:mismatch_PhD_HPhD_recal_LIGO}\includegraphics[width=0.489\textwidth]{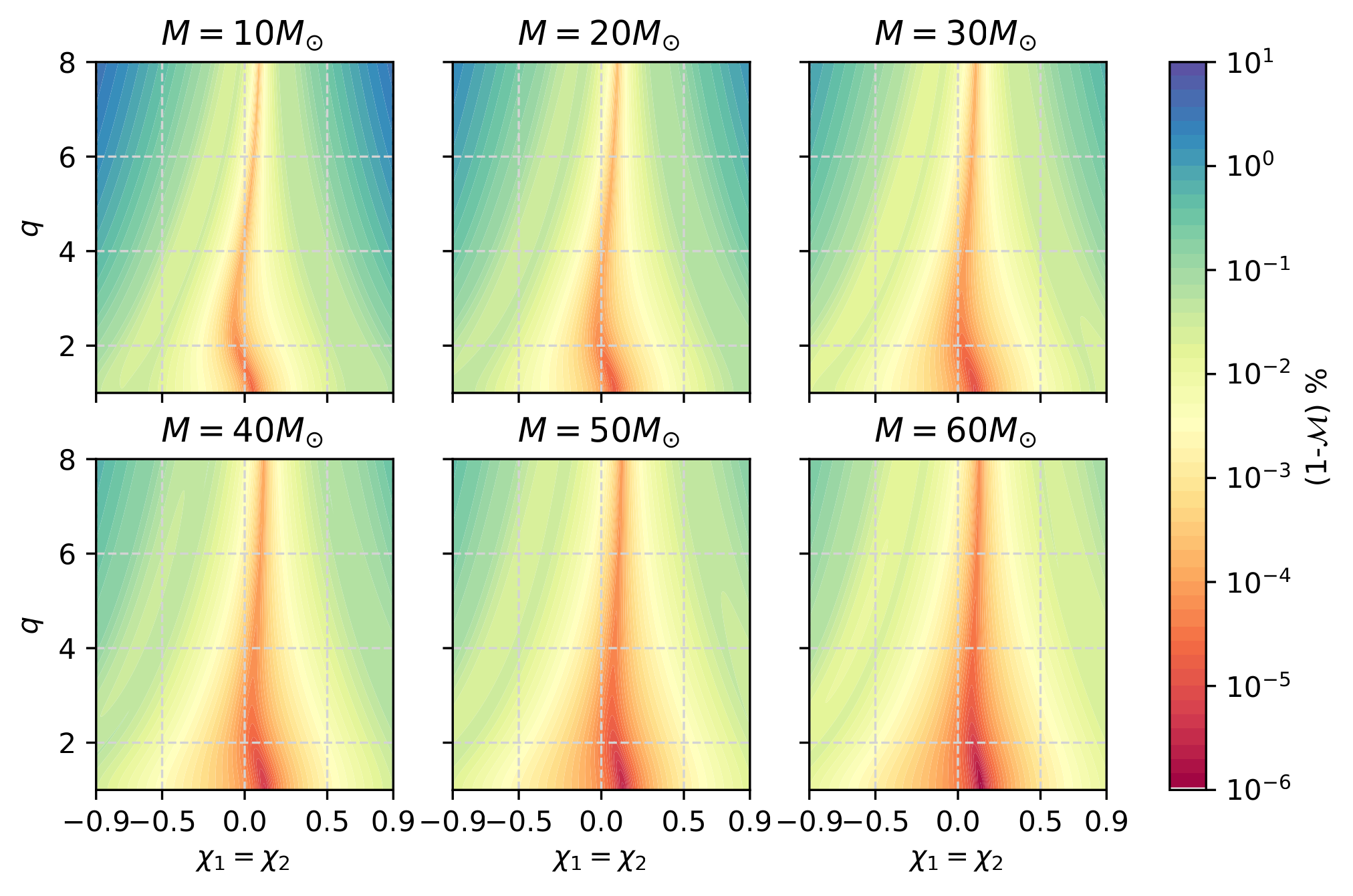}}
 \hfill
  \subfigure[]{\label{fig:mismatches_summary_1}\includegraphics[width=0.489\textwidth]{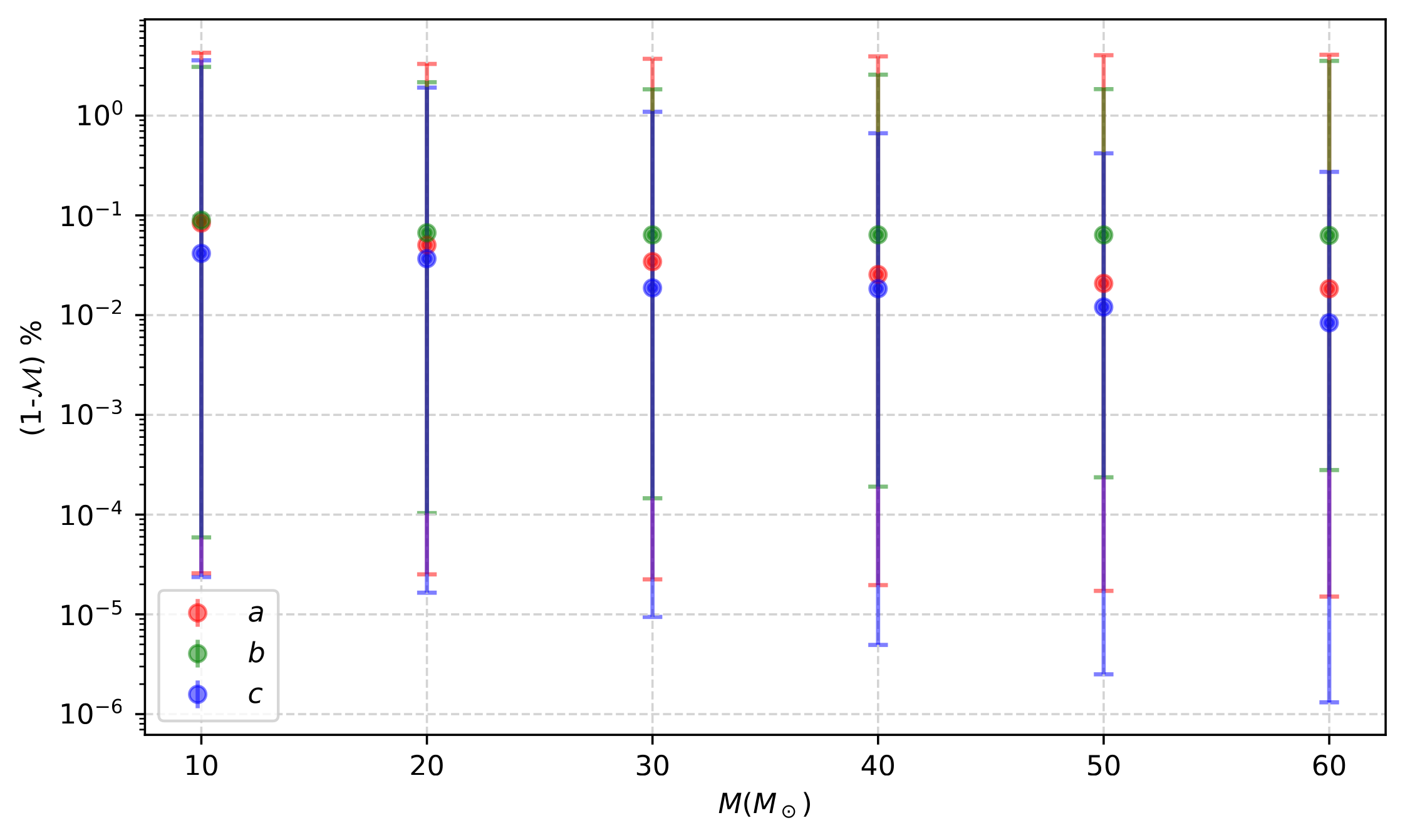}}
 \hfill
 \caption{ \ref{fig:mismatch_TF2_HTF2_LIGO}: Mismatches between 3.5PN \texttt{TaylorF2}, with the frequency-domain phase described in Appendix~\ref{app:tf2phase}, with and without the tidal heating phase correction derived in Ref.~\cite{Saketh:2022xjb}. A lower cutoff of 10Hz and an upper cutoff of $F=0.018$ are used, in the Advanced LIGO zero detuned high power noise PSD. \ref{fig:mismatch_TF2_HTF2_isco_LIGO}: Same as in Fig.~\ref{fig:mismatch_TF2_HTF2_LIGO}, but with an upper cutoff corresponding to the merger remnant KBH ISCO frequency described in Ref.~\cite{Favata:2021vhw}. \ref{fig:mismatch_PhD_HPhD_recal_LIGO}: Mismatches between \texttt{PhenomD\_Recal} (see Appendix~\ref{app: wf models}) and \texttt{PhenomD\_Horizon}. For both the models we followed the same steps, as outlined in this paper and in Refs.~\cite{Husa:2015iqa,Khan:2015jqa} -- the only difference being the presence of TH phase and amplitude corrections (for details, see Section~\ref{sec:comparison}). Using \texttt{PhenomD\_Recal} instead of the original \texttt{PhenomD} enables us to eliminate any modeling differences between the latter model and the present work. Cutoff frequencies for the match calculations are the same as in Fig.~\ref{fig:mismatch_TF2_HTF2_LIGO}.  In \ref{fig:mismatches_summary_1} we show the range of mismatches (in $\%$) for different binary masses. The red, green and blue colors represent the (a), (b) and (c) panels of the current figure, respectively. The top and bottom horizontal cap lines denote the maximum and minimum mismatch values respectively, and the solid circles correspond to the medians of the mismatch distribution at those mass values. For example, the red circle and lines at $M=10M_{\odot}$ represent the mismatch range at the upper-row-left-most subfigure of panel~\ref{fig:mismatch_TF2_HTF2_LIGO}. } 
 \label{fig:mismatch_TF2}
\end{figure*}

\begin{figure*}[btp]
\centering
 \subfigure[]{\label{fig:mismatch_D_phaseonly}\includegraphics[width=0.489\textwidth]{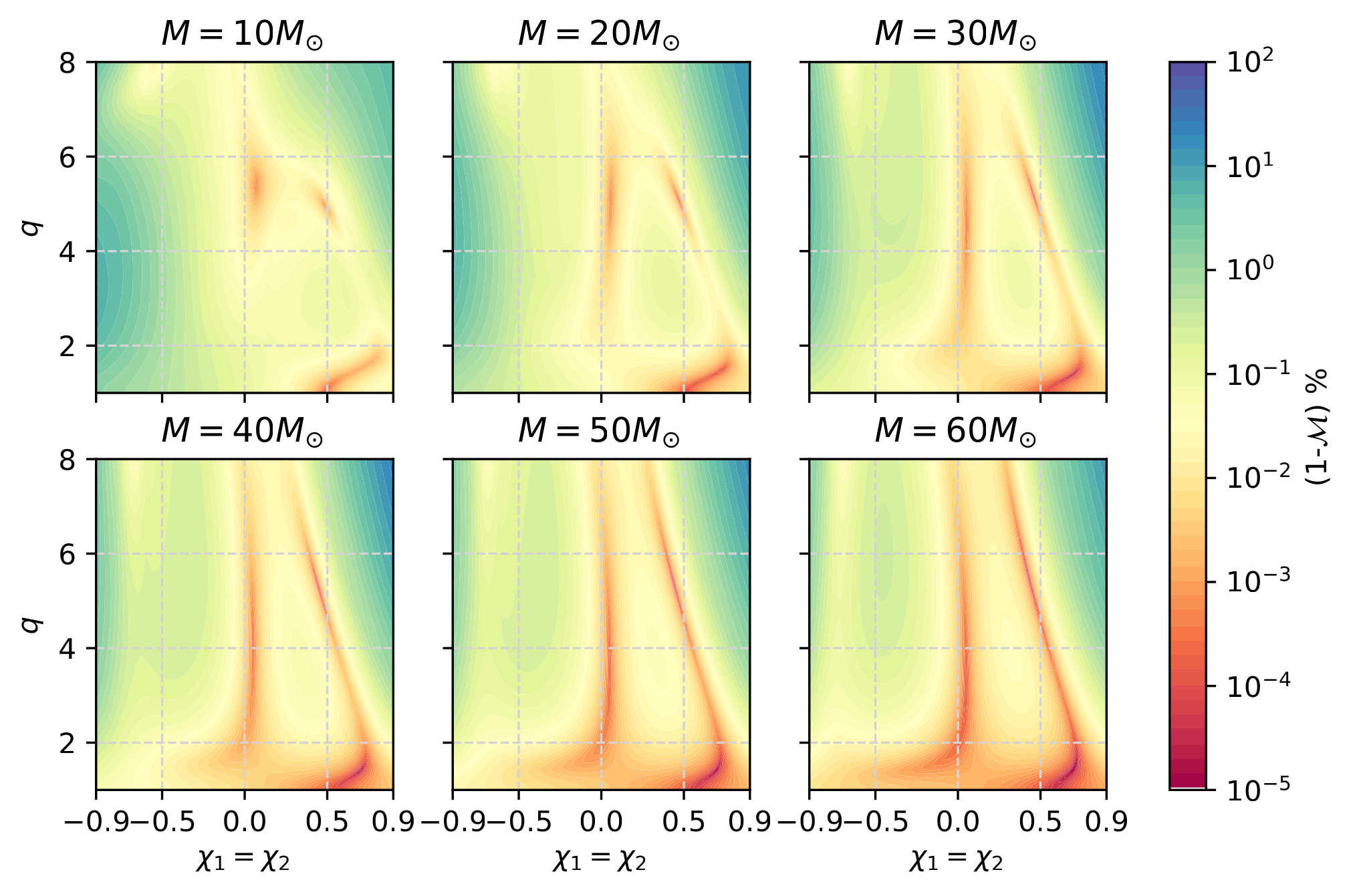}}
 \hfill
 \subfigure[]{\label{fig:mismatch_D_amponly}\includegraphics[width=0.489\textwidth]{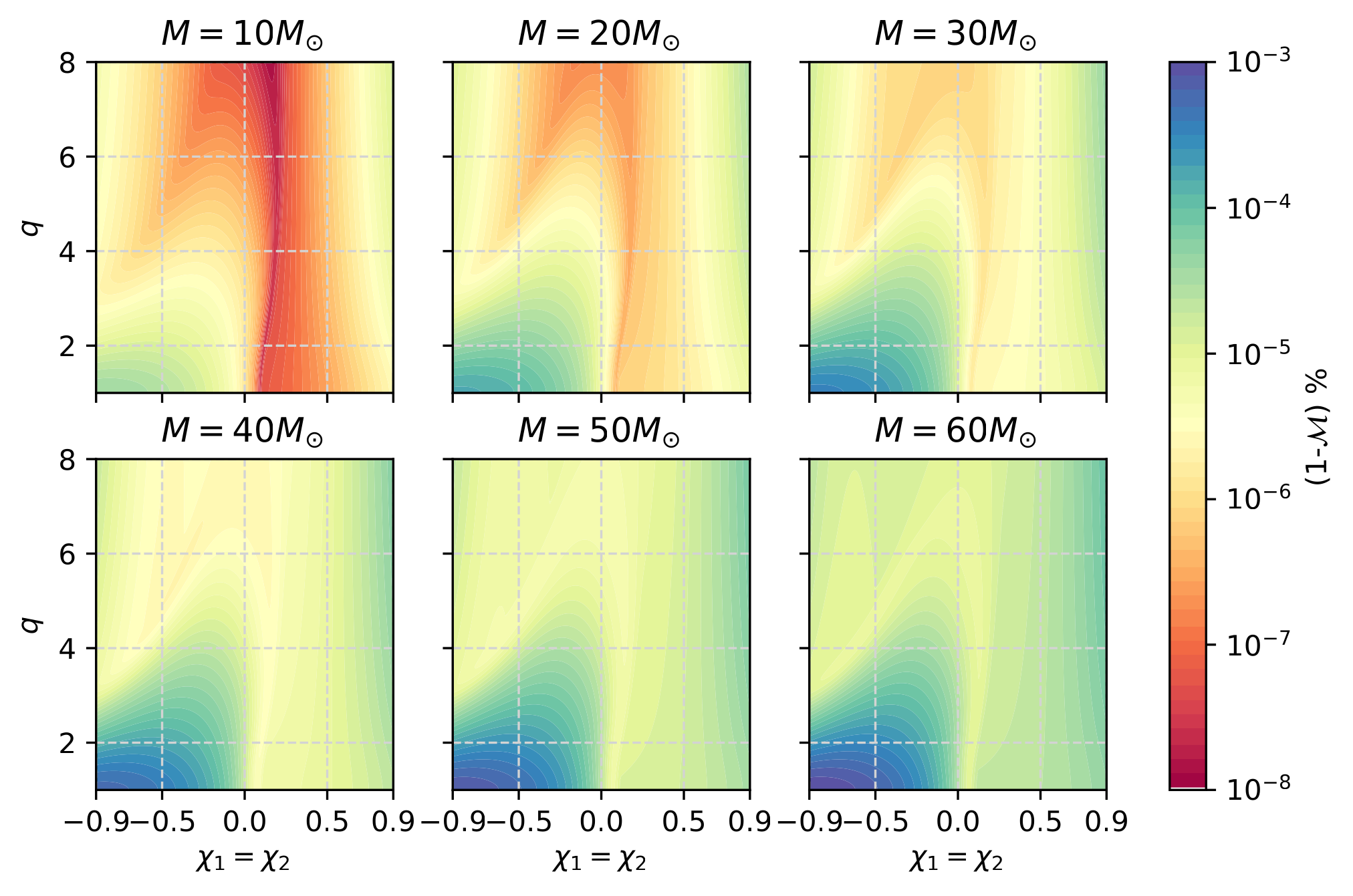}}
 \hfill
  \subfigure[]{\label{fig:mismatch_D_both}\includegraphics[width=0.489\textwidth]{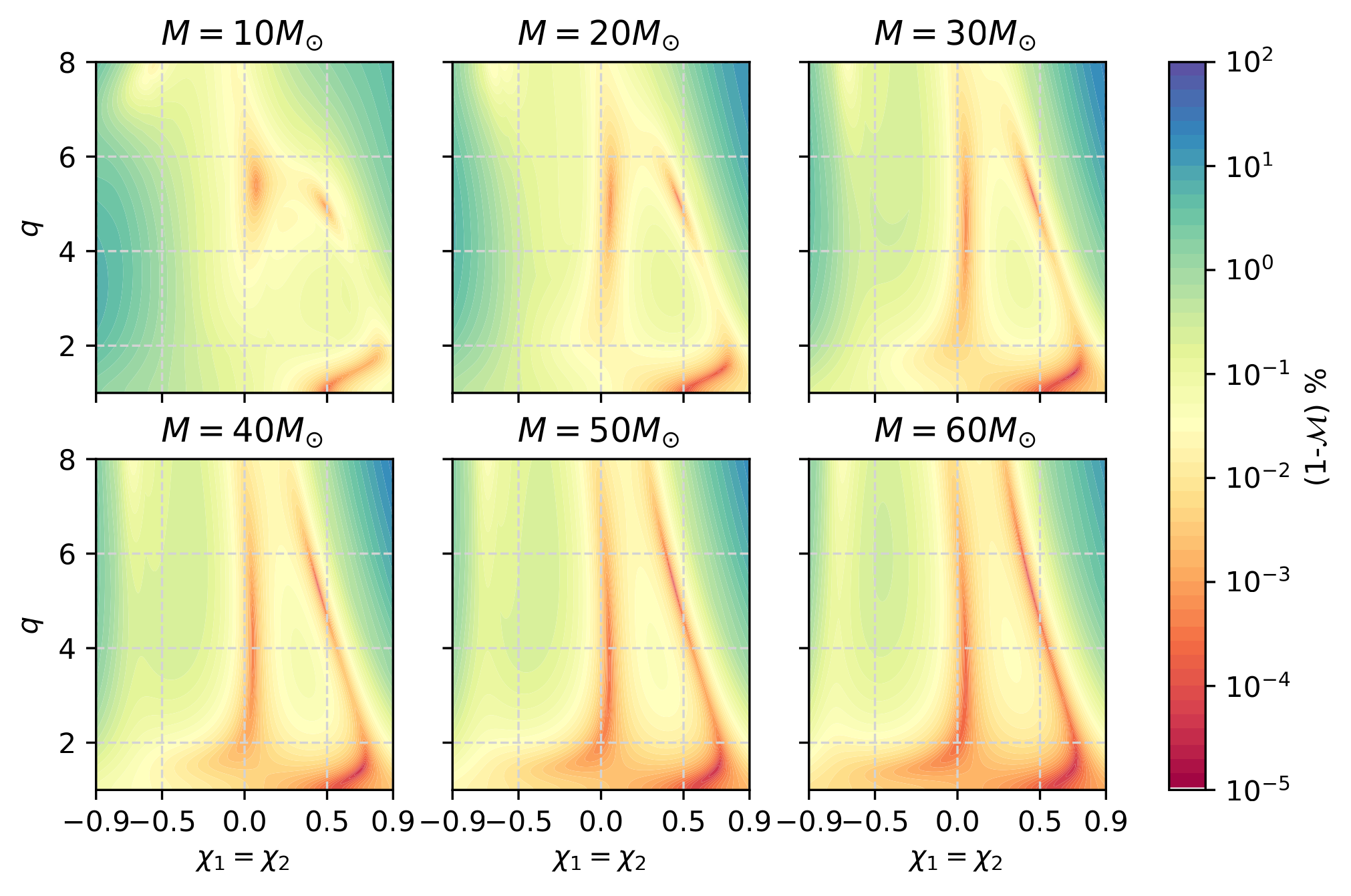}}
 \hfill
  \subfigure[]{\label{fig:mismatches_summary_2}\includegraphics[width=0.489\textwidth]{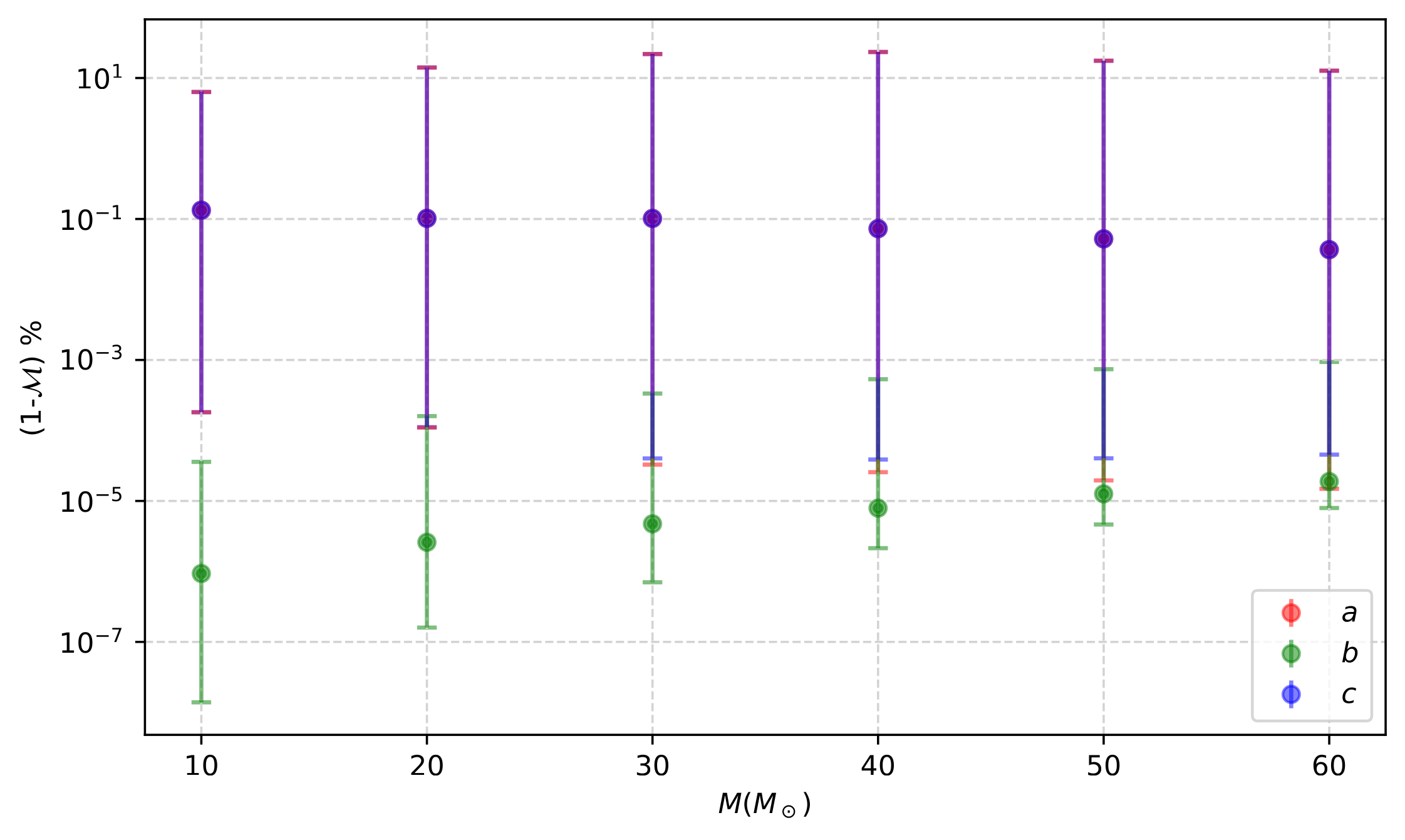}}
 \hfill
 \caption{Mismatches (\%) between \texttt{PhenomD\_Horizon} and \texttt{PhenomD}, with the lower frequency cutoff of $f_{\rm low}=10$ Hz. \ref{fig:mismatch_D_phaseonly}: Mismatches with only the phase correction but no amplitude correction in \texttt{PhenomD\_Horizon}. \ref{fig:mismatch_D_amponly}: Mismatches with only the amplitude correction without any phase correction. \ref{fig:mismatch_D_both}: Mismatches with corrections in both the phase and amplitude. The plots show that the dominant contributor to these mismatches is the phase correction. In~\ref{fig:mismatches_summary_2}, we show the range and median of mismatches (in $\%$) for different binary masses corresponding to the (a), (b) and (c) panels of the current figure, in a similar manner as in Fig.~\ref{fig:mismatches_summary_1}. }
 
\label{fig:mismatch_vs_D}
\end{figure*}

In Table~\ref{tab:hyb_nc} we list the extra test hybrids created for assessing the model accuracy. Figure~\ref{fig:mismatch} shows the mismatches between the entire set of test hybrids (Table~\ref{tab:hyb} and Table~\ref{tab:hyb_nc}) against the BBH waveform model presented in this paper. We have used the Advanced LIGO zero-detuned high power~\cite{KAGRA:2013rdx} (ZDHP) noise curve with a lower frequency cutoff of 10 Hz, and upper cutoff of $Mf=0.3$. The mismatches are plotted against the total binary mass. We find that the model accuracy is at par with the accuracy standard of \texttt{PhenomD}, with almost all the hybrids showing mismatches well below 1\%, the majority of them having mismatches around 0.1\% within the mass range $10M_\odot\leqslant M \leqslant 100M_\odot$. The worst mismatch we find is for the $\{q=4,\chi^{}_1=0,\chi^{}_2=0.8\}$ configuration, which crosses the 1\% level above $\sim 50M_\odot$, and reaches a maximum of 1.21\% at $70M_\odot$.

\section{Model comparison}\label{sec:comparison}

First, to look at how the TH corrections impact an analytical waveform model, we compare 3.5PN \texttt{TaylorF2}, with the frequency-domain phase defined in Appendix~\ref{app:tf2phase}, with and without the TH corrections in the phase and amplitude. Figure~\ref{fig:mismatch_TF2_HTF2_LIGO} shows the corresponding mismatches, keeping $\chi^{}_1=\chi^{}_2$. We explore the mismatches within the parameter space $q \in [1,8]$ and $\chi^{}_1,\chi^{}_2 \in [-0.9,0.9]$, for six values of the binary mass, $M/M_\odot \in [10,20,30,40,50,60]$. The frequency range for the mismatch calculation is $f\in [10,0.018/M]$ (Hz). Since the phase correction due to TH written in Eq.~\eqref{eq:phase correction} is $\propto 1/\eta$ at the leading order, higher mismatches are found for higher mass ratios, for a fixed value of the spins. With increasing spins, mismatches rise due to the increased significance of TH. Figure~\ref{fig:mismatch_TF2_HTF2_isco_LIGO} is similar to Fig.~\ref{fig:mismatch_TF2_HTF2_LIGO} in every way only except the upper cutoff frequency, which is chosen to be the ISCO of the corresponding merger remnant KBH~\cite{Favata:2021vhw}. Before we compare \texttt{PhenomD\_Horizon} and \texttt{PhenomD}, we note that there are some differences in modeling and calibrating the two waveforms apart from just the addition of the correction due to horizon fluxes. First, \texttt{PhenomD} was calibrated up to $q=18$, with a different set of hybrid waveforms. Second, the quadratic and cubic spin terms were absent in the \texttt{TaylorF2} phasing, which we take into account at the 3PN and 3.5PN orders, respectively. Thus, to discern the physical difference that \texttt{PhenomD\_Horizon} introduces, we create another model (\texttt{IMRPhenomD\_Recal}) by leaving out the TH corrections in the phase and amplitude, but keeping every other aspect unchanged. We expect this model, being free from any modeling differences with \texttt{PhenomD\_Horizon}, to produce mismatches comparable to Fig.~\ref{fig:mismatch_TF2_HTF2_LIGO}. We plot them in Fig.~\ref{fig:mismatch_PhD_HPhD_recal_LIGO} and, indeed, we see that the mismatches exhibit a qualitative similarity. However, since the upper cutoff is fixed at the inspiral cutoff of \texttt{PhenomD} ($Mf=0.018$), and the hybrids used for calibration of both \texttt{PhenomD\_Horizon} and \texttt{PhenomD\_Recal} are stitched to NR data before that, we see some differences between these two plots. Specifically, in Fig.~\ref{fig:mismatches_summary_1} we show the maxima, minima, and the medians of mismatches for each of the subplots in Figs.~\ref{fig:mismatch_TF2_HTF2_LIGO},~\ref{fig:mismatch_TF2_HTF2_isco_LIGO}, and~\ref{fig:mismatch_PhD_HPhD_recal_LIGO}. Overall, the analytical waveforms show higher mismatches than the phenomenological ones, which can be ascribed to the presence of NR data in the hybrids between the stitching frequency and the inspiral upper cutoff of $Mf=0.018$. With increasing binary mass, the differences are higher, which is expected since the similarity between \texttt{PhenomD\_Horizon} and \texttt{PhenomD\_Recal} are more towards higher frequencies.

Figure~\ref{fig:mismatch_vs_D} shows the mismatches between \texttt{IMRPhenomD\_Horizon} and the original \texttt{IMRPhenomD} \footnote{This should be distinguished from \texttt{PhenomD\_Recal} that we used for Fig.~\ref{fig:mismatch_PhD_HPhD_recal_LIGO}. We constructed \texttt{PhenomD\_Recal} to find out how much mismatch arises solely due to TH. In Fig.~\ref{fig:mismatch_vs_D}, however, the mismatch values with \texttt{PhenomD} originate both from modeling difference as well as TH. }, now in the frequency range $f\in [10,0.3/M]$ (Hz). In Fig.~\ref{fig:mismatch_D_phaseonly}, we show the mismatches including only the phase correction - i.e. using the recalibrated inspiral phase, but keeping the IMR amplitude the same as \texttt{PhenomD}. The mismatch reaches a maximum of $\sim 14\%$ for $30M_\odot$ binaries with $q\gtrsim 7$ and  $\chi^{}_1=\chi^{}_2\gtrsim 0.8$, and a minimum of $\sim 10^{-5}\%$ for heavier binaries. We see substantial regions of the parameter space with mismatches between 1-10\% for spin magnitudes of $|\chi^{}_1|=|\chi^{}_2|\gtrsim 0.5$, and $q\gtrsim 4$. 

Figure~\ref{fig:mismatch_D_amponly} shows mismatches when only the amplitude correction is added to the inspiral and the intermediate region is re-evaluated accordingly, but the phase is kept the same as \texttt{PhenomD}. The mismatches, in this case, are orders of magnitude less than the phase-corrected model, and never exceed 0.001\% within this parameter space. This result also vindicates our choice of not recalibrating the amplitude model, as discussed in Sec~\ref{sec:amplitude}. We also note from Eq.~\eqref{eq:amp correction} that at the leading order, $\Tilde{A}_{\rm TH}\propto \sqrt{\eta}$, which implies that the TH corrections are more significant for more symmetric masses. This is in contrast to the behaviour of the phase correction, and renders the amplitude correction ineffective for high mass-ratio binaries with notable TH contribution in the phase. Nevertheless, $\Tilde{A}^{}_{\rm TH}$ still rises with increasing spins. 

In Fig.~\ref{fig:mismatch_D_both}, we use the complete \texttt{IMRPhenomD\_Horizon} model, by including both the phase and amplitude corrections. The distribution of mismatches in this figure shows great similarity with Fig.~\ref{fig:mismatch_D_phaseonly}. In Fig.~\ref{fig:mismatches_summary_2}, we plot the maxima, minima, and medians of mismatches for all the subfigures, the same way we did for Fig.~\ref{fig:mismatches_summary_1}.  This suggests that including the amplitude correction in the TH corrected model has negligible effect compared to the phase correction in GW data analysis. 

\begin{figure}
\vspace{5mm}
    \centering
    \includegraphics[width=0.98\linewidth]{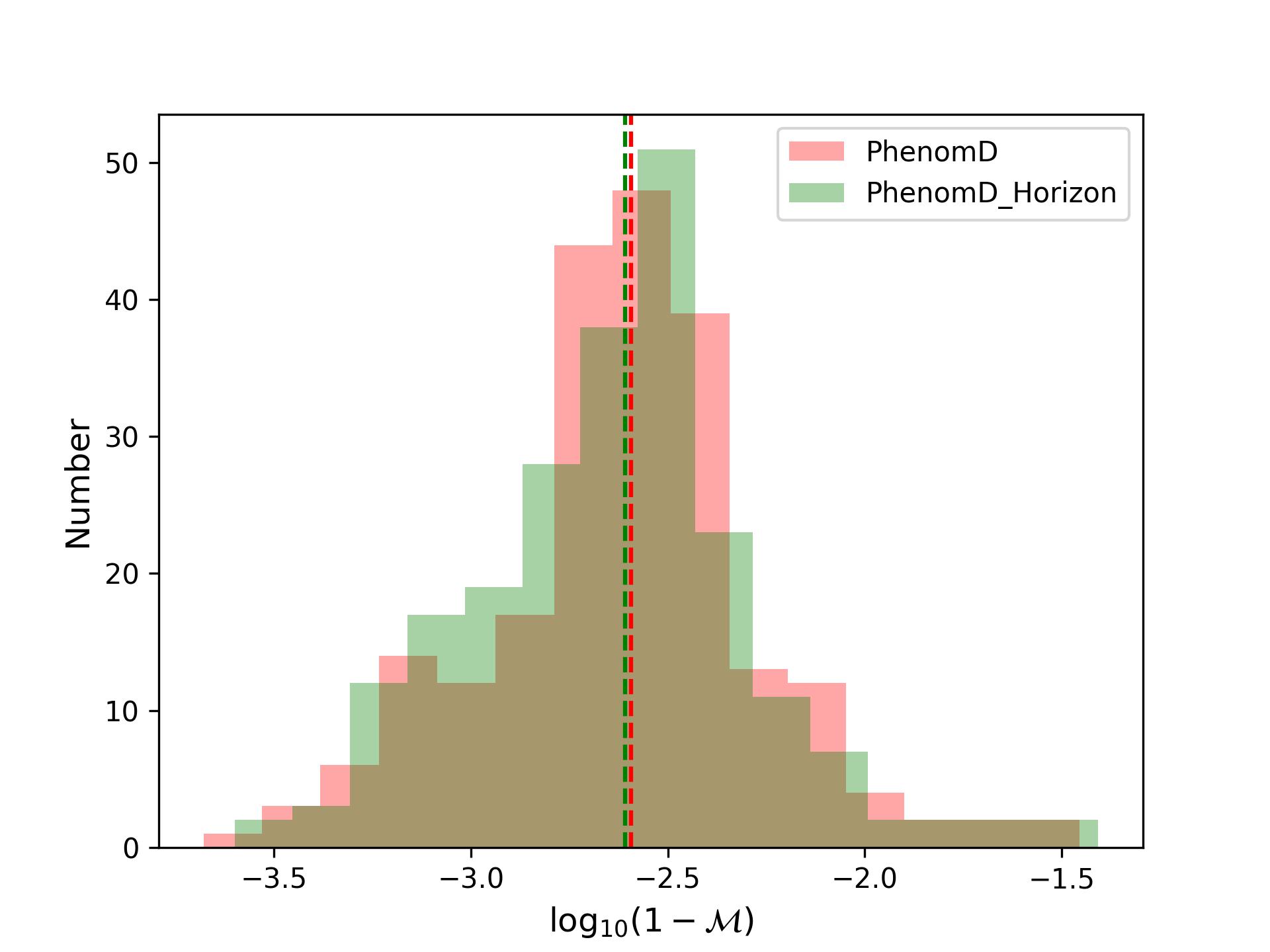}
    \caption{Mismatches of 219 non-precessing noneccentric NR waveforms from SXS with \texttt{PhenomD\_Horizon} and \texttt{PhenomD}, in aLIGO ZDHP noise curve, averaged in the binary mass range $12-100M_\odot$. The dashed lines show the median values of the distributions.}
    \label{fig:mismatchNRzdhp}
\end{figure}

\begin{figure}
\vspace{5mm}
    \centering
    \includegraphics[width=0.98\linewidth]{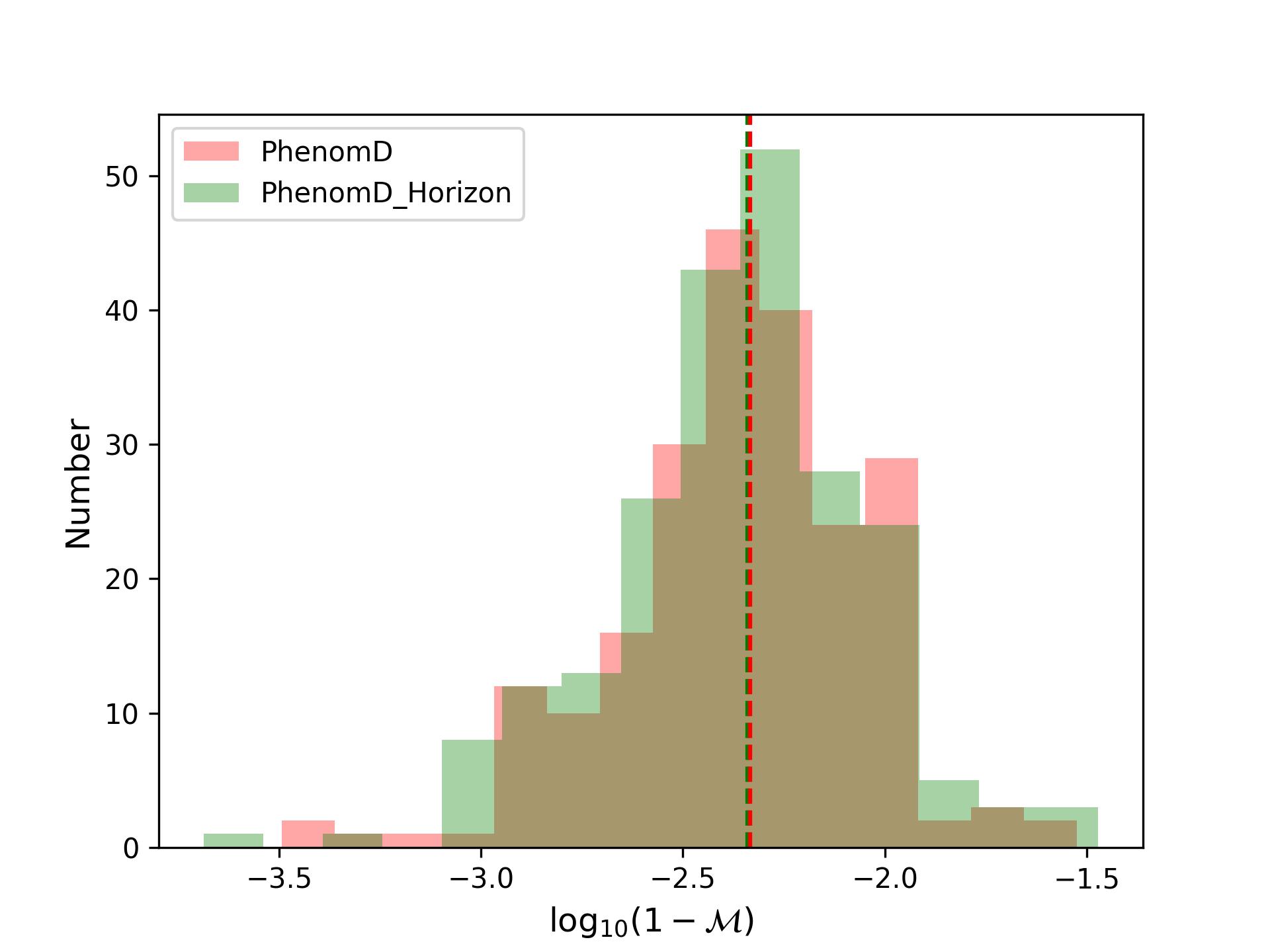}
    \caption{Same as in Fig.~\ref{fig:mismatchNRzdhp}, but with a flat noise curve.}
    \label{fig:mismatchNRflat}
\end{figure}

To demonstrate the improvement in accuracy that our model introduces over \texttt{PhenomD}, we use 219 NR waveforms from the SXS catalog and compare them against \texttt{PhenomD} and \texttt{PhenomD\_Horizon} separately by computing mismatches. In Fig.~\ref{fig:mismatchNRzdhp}, we show the distributions of the mismatches by using the Advanced LIGO ZDHP noise curve. We plot histograms of $\log_{10}(1-\mathcal{M})$, where $(1-\mathcal{M})$ is the average mismatch in the binary mass range $12M_\odot \leqslant M \leqslant 100M_\odot$. The dashed lines show the medians of the two histogram plots, where a slight overall improvement is seen. The fractional shift in the medians (denoted by an overbar), is
\begin{equation}
    \frac{\overline{(1-\mathcal{M})}_{\rm PhenomD} - \overline{(1-\mathcal{M})}_{\rm PhenomD\_Horizon}}{\overline{(1-\mathcal{M})}_{\rm PhenomD}} \approx 0.04,
\end{equation}
showing an improvement of $\sim 4\%$. Figure~\ref{fig:mismatchNRflat} shows similar histograms with a flat noise curve. In this case, we also see a shift in the median value towards lower mismatches, with an improvement of $\sim 1.5\%$.


\section{Discussion and future work}\label{sec:discussion}

The main objective of this work is to present a phenomenological gravitational waveform model \texttt{IMRPhenomD\_Horizon} for binary black holes that carries the signatures of tidal heating of the holes explicitly throughout the binary evolution, starting from the early inspiral phase and ending at their merger. We also demonstrate, in Sec~\ref{sec:phase comparison}, that inclusion of corrections due to tidal heating of black holes, in the phase and amplitude, rewards an overall improvement in agreement between the point-particle analytical description (here it is uncalibrated \texttt{SEOBNRv2}) of a compact binary coalescence and numerical relativity data, especially for high mass-ratios and high spins. In the frequency domain, the waveform model presented here contains corrections due to this effect both in the phase and amplitude. The entire IMR model for the phase is divided into two parts. The inspiral phase model, within the frequency range $0.0035 \leqslant Mf\leqslant 0.018$, is a recalibrated version of \texttt{IMRPhenomD}~\cite{Husa:2015iqa,Khan:2015jqa} to account for the modifications due to tidal heating. The merger-ringdown parts, defined as $Mf\geqslant 0.018$ in terms of the dimensionless frequency, are the same as \texttt{PhenomD} since this region contains purely NR data, which inherently incorporates the dephasing due to tidal heating. The amplitude corrections are added by dividing the model into three parts. The inspiral model, with $0.0035 \leqslant Mf\leqslant 0.014$, is constructed by directly adding the correction term to the amplitude model of \texttt{PhenomD} in this region. The intermediate region, $0.014\leqslant Mf \leqslant Mf_{\rm peak}$, $f_{\rm peak}$ being the frequency corresponding to the peak amplitude, is approximated as a polynomial in $f$ and evaluated by imposing $C^{(1)}$ continuity at the two ends and a collocation point at their mid-frequency. The amplitude model is identical to \texttt{PhenomD} for $f>f_{\rm peak}$. The final model is calibrated within the range in the mass ratio $q\in [1,8]$ and the effective spin parameter $\chi^{}_{\rm PN}\in [-0.79,0.72]$.

We conducted a study of faithfulness by calculating mismatches between the model and a validation dataset of 40 hybrid waveforms constructed with tidal heating included. We demonstrated that modeling errors do not lead to mismatches higher than 1\% (barring a single outlier), with most of the mismatch values lying around 0.1\%. This standard of faithfulness is comparable to \texttt{PhenomD}, which also considered a maximum tolerance level of 1\% mismatch.

How the new model differs from its parent model \texttt{PhenomD} was assessed by calculating mismatches between them within the calibration range of the intrinsic parameter space, for binary masses ranging from $10M_\odot$ to $60M_\odot$. Phase corrections in the inspiral result in mismatches between these two models as high as $\sim 14\%$, whereas amplitude corrections alone are unable to produce mismatches above $\sim 0.001\%$. With increasing mass asymmetry, the impact of the phase correction increases, but the amplitude correction becomes less effective. Combined with the fact that the former has more significance in GW searches and PE, this reciprocal nature of these two correction terms leaves the latter mostly superfluous for practical purposes. 

We compared the model with a set of 219 NR waveforms from the SXS catalog~\cite{Boyle:2019kee}, and show the mismatch distribution along with the same for \texttt{PhenomD}, averaged over the binary mass range $12M_\odot \leqslant M \leqslant 100M_\odot$. With the Advanced LIGO ZDHP noise curve, we see an improvement of $\sim 4\%$ in the medians of the two distributions. With a flat noise curve, the improvement is relatively modest, about 1.5\%. These overall improvements point towards a mild yet non-negligible increase in waveform accuracy. 

Tidal heating of black holes, after all, is a weak effect in the context of comparable-mass binaries. Nevertheless, an absence of its signatures from the complete waveform models can 
bias the estimated parameters of a binary. Its absence can also mimic a deviation from GR predictions, raising complications in tests of GR. We leave these studies to future work. Additionally, tidal heating has the important property of varying significantly according to the nature of the component objects, which makes it a potential discriminator for BHs from other objects. However, to utilize the full power of TH as a BH identifier, one needs to model the energy and angular momentum fluxes down the BH horizons also in the late inspiral regime described by NR, where the tidal fields are the strongest. The feasibility and effectiveness of modeling these quantities can be explored in future studies, where the model presented in this work can be used directly.

The current work, in its entirety, is based on the frequency-domain waveform model \texttt{IMRPhenomD}. For this reason, we perform all the comparisons of the model presented here with the said approximant only. More recent and state-of-the-art phenomenological waveform models are available currently, namely, the \texttt{IMRPhenomX}~\cite{Pratten:2020fqn,Pratten:2020ceb} family of waveforms. These waveforms are calibrated with a much larger set of hybrid waveforms, and use superior techniques like hierarchical modeling~\cite{Jimenez-Forteza:2016oae} to reduce modeling errors, and also to generate more accurate waveforms with unequal spins. These models boast 1 to 2 orders of magnitude better faithfulness than \texttt{PhenomD}, and a dramatic shift in the mismatch distribution against NR waveforms towards lower values~\cite{Pratten:2020fqn}. Including the effects of TH in these models by recalibrating the inspiral is beyond the scope of this work, which uses only publicly available NR data from the SXS catalog. We leave them as future possibilities.

\section*{Acknowledgments}
SM thanks the University Grants Commission (UGC), India, for providing financial support as a senior research fellowship. KSP gratefully acknowledges the support of the Science and Technology Facilities Council (STFC) of the United Kingdom under the grant ST/V005677/1. The high performance computing facilities \texttt{Pegasus} at IUCAA  and {\tt Bluebear} at the University of Birmingham were used for this work.  The authors are grateful for computational resources provided by the LIGO Lab and supported by National Science Foundation Grants PHY-0757058 and PHY-0823459. The authors thank Vaishak Prasad, Apratim Ganguly, Anuj Mishra, Mark Scheel, Nathan Johnson-McDaniel, P. Ajith, and Shrobana Ghosh for useful comments and suggestions. The Python software packages \texttt{PyCBC}~\cite{alex_nitz_2023_7547919}, \texttt{Scipy}~\cite{Virtanen:2019joe}, \texttt{Numpy}~\cite{Harris:2020xlr}, \texttt{Matplotlib}~\cite{hunter} were used for computational works, analyses, and plots. \texttt{Mathematica}~\cite{Mathematica} was used for data fitting purposes. SB acknowledges the hospitality of IUCAA with gratitude and support from the NSF under Grant PHY-2309352.
\onecolumngrid

\appendix

\section{Post-Newtonian inspiral phase}\label{app:tf2phase}

We write the 3.5PN frequency-domain phase (without TH) as
\begin{equation}
    \Psi^{}_{\rm TF2}(f) = 2\pi f t^{}_c - \phi^{}_c -\frac{\pi}{4} + \frac{3}{128\eta v^5}\sum_{n=0}^7 \psi^{}_{(n)} v^n\,,
\end{equation}
where $v=(\pi Mf)^{1/3}$.

The coefficients $\{\psi^{}_{(n)}\}$ ($n$=0-7) are given by~\cite{Arun:2008kb,Mishra:2016whh,Wade:2013hoa},
\begin{align}
    \psi^{}_{(0)} =~& 1\,,\\
    \psi^{}_{(1)} =~& 0\,,\\
    \psi^{}_{(2)} =~& \frac{3715}{756} + \frac{55}{9}\eta\,,\\
    \psi^{}_{(3)} =~& -16\pi + \frac{113}{3}\delta\chi^{}_a + \left(\frac{113}{3} - \frac{76}{3}\eta\right)\chi^{}_s\,,\\
    \psi^{}_{(4)} =~& \frac{15293365}{508032} + \frac{27145}{504}\eta+ \frac{3085}{72}\eta^2 - \frac{405}{4}\delta\chi^{}_a\chi^{}_s + \left(-\frac{405}{8} + 200\eta\right)\chi_a^2 + \left(-\frac{405}{8} + \frac{5}{2}\eta\right)\chi_s^2\,, \\
    \psi^{}_{(5)} =~& \left[\frac{38645\pi}{756} - \frac{65\pi}{9}\eta - \left(\frac{732985}{2268} - \frac{24260}{81}\eta - \frac{340}{9}\eta^2\right)\chi^{}_s - \left(\frac{732985}{2268} + \frac{140}{9}\eta\right)\delta\chi_a\right]\left(1+3\ln{v}\right)\,,\\
    \psi^{}_{(6)} =~& \frac{11583231236531}{4694215680} - \frac{640\pi^2}{3} - \frac{6848}{21}\gamma^{}_E + \eta\left(-\frac{15737765635}{3048192} + \frac{2255\pi^2}{12}\right) + \frac{76055}{1728}\eta^2 - \frac{127825}{1296}\eta^3
         - \frac{6848}{21}\ln{(4v)}\nonumber \\&  + \pi\left\{\frac{2270}{3}\delta\chi^{}_a + \left(\frac{2270}{3} - 520\eta\right)\chi^{}_s\right\} + \left(\frac{75515}{144} - \frac{8225}{18}\eta\right)\delta\chi^{}_a\chi^{}_s 
         + \left(\frac{75515}{288} - \frac{263245}{252}\eta - 480\eta^2\right)\chi_a^2 \nonumber \\&  + \left(\frac{75515}{288} - \frac{232415}{504}\eta + \frac{1255}{9}\eta^2\right)\chi_s^2\,, \\
    \psi^{}_{(7)} =~& \frac{77096675\pi}{254016} + \frac{378515\pi}{1512}\eta- \frac{74045\pi}{756}\eta^2 + \left\{-\frac{25150083775}{3048192} + \frac{10566655595}{762048}\eta - \frac{1042165}{3024}\eta^2 + \frac{5345}{36}\eta^3\right. \nonumber \\&
         + \left.\left(\frac{14585}{8} - 7270\eta + 80\eta^2\right)\chi_a^2\right\}\chi^{}_s + \left\{\left(\frac{-25150083775}{3048192}
         + \frac{26804935}{6048}\eta - \frac{1985}{48}\eta^2\right)\chi^{}_a\right. \nonumber \\& \left. + \left(\frac{14585}{24} - 2380\eta\right)\chi_a^3 + \left(\frac{14585}{8} - \frac{215}{2}\eta\right)\chi^{}_a\chi_s^2\right\}\delta + \left(\frac{14585}{24} - \frac{475}{6}\eta + \frac{100}{3}\eta^2\right)\chi_s^3 \,.
\end{align}
Here $\delta = \sqrt{1-4\eta}$, $\chi^{}_s=(\chi^{}_1+\chi^{}_2)/2$, $\chi^{}_a=(\chi^{}_1-\chi^{}_2)/2$. $\gamma^{}_E$ is the Euler's constant, $\gamma^{}_E=0.5772156\cdots$.

\section{Nomenclature of various waveform models (existing or temporarily introduced)}\label{app: wf models}

\begin{table}[h]
    \centering
    \begin{tabular}{c|c}
        \hline
        Name of the model & Description \\
        \hline
        \hline\\
        \texttt{SEOBNRv2} & The EOB waveform of the same name as published in Ref.~\cite{Taracchini:2013rva}.\\
        \hline\\
        \texttt{SEOBv2} & The EOB model after removing the NR calibratios from \texttt{SEOBNRv2}.\\
        \hline\\
        \texttt{SEOBv2\_Horizon} & The time domain model after adding the corrections due to horizon fluxes\\ & of black holes to the phase and the amplitude of \texttt{SEOBv2}.\\
         \hline\\
         \texttt{IMRPhenomD}\\ (abbreviated as \texttt{PhenomD}) & The phenomenological IMR waveform of the same name as published in Refs.~\cite{Husa:2015iqa,Khan:2015jqa}.\\
         \hline\\
         \texttt{IMRPhenomD\_Horizon} & The final frequency-domain IMR model of BBHs with the horizon flux\\ & corrections in phase and amplitude, the main objective of this work.\\
        \hline\\
        \texttt{IMRPhenomD\_Recal} & A recalibrated version of \texttt{IMRPhenomD} we created for comparing with\\ &  \texttt{IMRPhenomD\_Horizon}, since there are differences in the calibration range and the\\ & \texttt{TaylorF2} phase that have been used in the former model and the current work.\\ & For this, we use \texttt{SEOBv2} hybrids instead of \texttt{SEOBv2\_Horizon}, and leave out\\ & $\Psi_{\rm TH}$ in Eq.~\eqref{eq:phiins}.\\
        \hline

    \end{tabular}
    \caption{}
    \label{tab:wf models}
\end{table}

\section{Ringdown and peak frequency}\label{app:ringdown}

The quasi-normal mode (QNM) frequency of a BH can be expressed in terms of its oscillatory (real) and damping (imaginary) parts as,
\begin{equation}
    f^{}_{\rm QNM} = f^{}_{\rm RD} - if^{}_{\rm damp}\,.
\end{equation}
$f_{\rm RD}$ can be obtained from the fitting formula~\cite{Nakano:2018vay}
\begin{equation}
    f^{}_{\rm RD} = \frac{1}{2\pi M_f}\left[f^{}_1+f^{}_2(1-\chi^{}_f)^{f_3}\right]\,,
\end{equation}
where $M^{}_f$ and $\chi^{}_f$ are the mass and spin of the merger remnant BH. Evaluation of these quantities is described in Appendix C of Ref~\cite{Favata:2021vhw}.
For the dominant harmonic mode $l=m=2$, the coefficients are given by, $f^{}_1=1.5251,\,f^{}_2=-1.1568,\,f^{}_3=0.1292$.

$f^{}_{\rm damp}$ can be written as
\begin{equation}
    f^{}_{\rm damp} = \frac{f^{}_{\rm RD}}{2Q},
\end{equation}
where the quality factor $Q$ can be expressed by the fitting formula
\begin{equation}
    Q = q^{}_1+q^{}_2(1-\chi^{}_f)^{q_3}.
\end{equation}
For the $l=m=2$ mode, $q^{}_1=0.7,\,q^{}_2=1.4187,\,q^{}_3=-0.499$.

The frequency corresponding to the peak amplitude in the Fourier domain, $f_{\rm peak}$, can then be obtained from the phenomenological relation
\begin{equation}
    f^{}_{\rm peak} = \left| f^{}_{\rm RD} + \frac{f^{}_{\rm damp}\gamma^{}_3(\sqrt{1-\gamma_2^2}-1)}{\gamma^{}_2}  \right|\,,
\end{equation}
$\gamma^{}_2$ and $\gamma^{}_3$ being two phenomenological parameters that can be evaluated for given values of $\{\eta,\chi^{}_1,\chi^{}_2\}$ from the ansatz in Eq.~\eqref{eq:Lambda}. The corresponding coefficients are given in Appendix C of Ref.~\cite{Khan:2015jqa}.

\section{Coefficients for the pseudo-PN parameters}\label{app:lambda}

\begin{table}[h]
    \centering
    {\renewcommand{\arraystretch}{1.6}
    \begin{tabular}{p{0.06\linewidth}|p{0.18\linewidth}|p{0.22\linewidth}|p{0.22\linewidth}|p{0.22\linewidth}}
        \hline
         & $\sigma^{}_1$ & $\sigma^{}_2$ & $\sigma^{}_3$ & $\sigma^{}_4$ \\
        \hline
        \hline
        $\lambda^{}_{00}$ & 730.759 & 25127.527 & $-$130492.02 & 203261.985 \\\hline
        $\lambda^{}_{10}$ & 2043.816 & $-$146822.249 & 714488.221 & $-$1.0914$\times 10^6$ \\\hline
        $\lambda^{}_{01}$ & $-$5997.126 & 181096.89 & $-$809867.9 & 1.1859$\times 10^6$ \\\hline
        $\lambda^{}_{11}$ & 54465.793 & $-$1.48114$\times 10^6$ & 6.566$\times10^6$ & $-$9.5763$\times10^6$ \\\hline
        $\lambda^{}_{21}$ & $-$88292.028 & 2.9436$\times10^6$ & $-$1.275$\times10^7$ & 1.8316$\times10^7$ \\\hline
        $\lambda^{}_{02}$ & $-$12729.742 & 249044.619 & $-$1.0889$\times 10^6$ & 1.5653$\times10^6$ \\\hline
        $\lambda^{}_{12}$ & 118443.267 & $-$2.3723$\times10^6$ & 1.0356$\times10^7$ & $-$1.4902$\times10^7$ \\\hline
        $\lambda^{}_{22}$ & $-$275615.32 & 5.7947$\times10^6$ & $-$2.51$\times 10^7$ & 3.6013$\times10^7$ \\\hline
        $\lambda^{}_{03}$ & $-$4715.744 & 86494.178 & $-$373232.315 & 524127.359 \\\hline
        $\lambda^{}_{13}$ & 52582.588 & $-$947035.327 & 4.0911$\times10^6$ & $-$5.8335$\times10^6$ \\\hline
        $\lambda^{}_{23}$ & $-$139301.918 & 2.5658$\times10^6$ & $-$1.1108$\times10^7$ & 1.5942$\times10^7$ \\
        
        \hline
    \end{tabular}}
    \caption{Best-fit values of $\lambda^i_{jk}$ parameters in Eq.~\eqref{eq:Lambda} for the pseudo-PN parameters $\{\sigma^{}_i\}$ in the inspiral phase.}
    \label{tab:lambda}
\end{table}

\FloatBarrier

\twocolumngrid

\bibliography{references}
\end{document}